\documentclass[prx,twocolumn,superscriptaddress,amsmath,amssymb]{revtex4-2}

\usepackage{graphicx}
\usepackage{dcolumn}
\usepackage{bm}

\usepackage[section]{placeins} 

\usepackage[utf8]{inputenc}
\usepackage[T1]{fontenc}
\usepackage{mathptmx}
\usepackage{etoolbox}
\usepackage{physics}
\usepackage{soul}
\usepackage{color}
\usepackage{comment}

\usepackage{hyperref}
\hypersetup{colorlinks=true, citecolor=blue, urlcolor=blue, linkcolor=blue}

\makeatletter
\def\@email#1#2{%
 \endgroup
 \patchcmd{\titleblock@produce}
  {\frontmatter@RRAPformat}
  {\frontmatter@RRAPformat{\produce@RRAP{*#1\href{mailto:#2}{#2}}}\frontmatter@RRAPformat}
  {}{}
}%
\makeatother

\renewcommand{\selectlanguage}[1]{}

\DeclareUnicodeCharacter{2212}{-}
\DeclareUnicodeCharacter{03B1}{$\alpha$} 
\DeclareUnicodeCharacter{0394}{$\Delta$} 

\begin{document}

\preprint{preprint number}

\title{Coherent microwave control of coupled electron-muon centers}
\author{Andrin Doll}
  \email{andrin.doll@psi.ch.}
\affiliation{PSI Center for Neutron and Muon Sciences CNM, Villigen PSI, Switzerland}%
\affiliation{PSI Center for Photon Sciences CPS, Villigen PSI, Switzerland}%

\author{Chennan Wang}
\affiliation{PSI Center for Neutron and Muon Sciences CNM, Villigen PSI, Switzerland}%

\author{Thomas Prokscha}
\affiliation{PSI Center for Neutron and Muon Sciences CNM, Villigen PSI, Switzerland}%

\author{Jan Dreiser}
\affiliation{PSI Center for Photon Sciences CPS, Villigen PSI, Switzerland}%

\author{Zaher Salman}
 \email{zaher.salman@psi.ch.}
\affiliation{PSI Center for Neutron and Muon Sciences CNM, Villigen PSI, Switzerland}%

\date{\today}

\begin{abstract}
Coherent control by means of tailored excitation is a key to versatile experimental schemes for spectroscopic investigation and technological utilization of quantum systems. Here we study a quantum system which consists of a coupled electron-moun spin state, i.e., muonium, a light isotope of hydrogen.
We demonstrate the most fundamental coherent control techniques by microwave excitation of spin transitions, namely driven Rabi oscillations and Ramsey fringes upon free evolution. Unprecedented performance is achieved by the microwave hardware devised for these experiments, which enables coherent spin manipulation of individual, isolated, muonium centers. For muonium formed in SiO$_2$ with strong electron-muon hyperfine interaction, a virtually undamped free precession signal is observed up to a 3.5 \textmu{}s time window. For muonium formed in Si with weak and anisotropic hyperfine interaction, a strong drive at the multi-quantum transition decouples the muonium center from its magnetic environment formed by the bath of $^{29}$Si nuclear spins at natural abundance. We expect that these capabilities will provide a powerful tool to investigate the effect of the environment on isolated coupled spins, uncover the details of coupled electron-muon systems in matter and validate quantum electrodynamics in the context of muonium spectroscopy. 

\end{abstract}

\maketitle

\section{\label{sec:intro}Introduction}

Controlling the state of a quantum system is essential for elaborate spectroscopic characterization and for emerging quantum technologies and applications. One such application is quantum sensing, which employs a quantum system to measure a physical quantity that couples to the sensor \cite{degen_quantum_2017}. A versatile approach to quantum sensing is muon spin rotation (\textmu{}SR) spectroscopy, where the spin of implanted muons probes its local magnetic environment. The principal observable in \textmu{}SR is the muon decay asymmetry, which corresponds to the spin dynamics of muons implanted into the sample with a well-defined spin polarization. Technically, the muon decay asymmetry is detected via the spatio-temporal emission of positrons that are released when the positively charged \mbox{(anti-)moun} with lifetime of 2.2 \textmu{}s decays. \textmu{}SR is therefore a technique that combines spin dynamics as encountered in related spectroscopic magnetic resonance techniques with exceptional single-spin sensitivity thanks to particle detection. Since muons can be implanted into a wide range of samples under various conditions, \textmu{}SR contributes to a diverse range of scientific problems ranging from fundamental to applied sciences \cite{hillier_muon_2022}.

Unlike experimental protocols in nuclear magnetic resonance (NMR) or electron paramagnetic resonance (ESR) that rely extensively on resonant manipulation of spin transitions, resonant spin excitation is less prominent in \textmu{}SR. In essence, coherent precessional spin dynamics are accessed in the transverse field (TF) geometry, where the spin of incoming muons is transverse to the externally applied field. The longitudinal muon spin polarization, including longitudinal relaxation $T_1$, is observed when implanting muons with spin along the external field in the longitudinal field geometry (LF). Furthermore, \textmu{}SR provides unique insight into static and dynamic magnetism at zero applied field, which is pivotal for the study of magnetic order in condensed matter. Thus, a variety of fundamental magnetic properties can be obtained without resonant excitation of the muon spin with its gyromagnetic ratio $\gamma_\mu/2\pi$ = 135.5 MHz/T.

In insulators, semiconductors, and soft matter, the positive charge of the muon is weakly screened. Accordingly, the muon might bind with an electron to form muonium, a light isotope of hydrogen \cite{patterson_muonium_1988,mckenzie_positive_2013}. In these coupled systems, transition frequencies often extend into the microwave frequency regime due to the larger gyromagnetic ratio of the electron spin as well as the electron-muon hyperfine coupling that can be as large as 4.5 GHz in vacuum muonium. The larger transition frequencies render resonant excitation beneficial under certain circumstances, as exemplied in the following. 

First, the direct time-domain observation of gigahertz precession frequencies remains a challenge until today \cite{stoykov_high-field_2012}, such that precision spectroscopy of vacuum muonium to test fundamental physics relies on microwave excitation \cite{favart_precision_1971,thompson_muonium_1973,casperson_new_1975,liu_high_1999,nishimura_rabi-oscillation_2021,mu-mass_collaboration_precision_2022, strasser_precision_2025}. Second, muonium centers with weak hyperfine interaction might feature \textmu{}SR-silent ESR transitions that can be probed by double electron-muon resonance (DEMUR) techniques to provide information about the coupled electron's orbital via the electron $g$-factor \cite{brown_detection_1979,blazey_double_1986,lord_double-resonance_2004}. Third, muonium formation might be delayed with respect to muon implantation, which causes line broadening in ordinary \textmu{}SR spectra. Such delayed formation products can be studied with resonant excitation to obtain insight into chemical reactions \cite{morozumi_muonium_1986, kreitzman_rf_1991, scheuermann_radio-frequency_1997, cottrell_radio-frequency_2003,johnson_muon_2004, mckenzie_hyperfine_2013,gfleming_rate_2015, dehn_direct_2020}. Previous DEMUR and delayed formation experiments employed radio-frequency (RF) excitation in the MHz regime and efforts to extend these techniques to the microwave regime remained singular \cite{kreitzman_microwave_1994,lord_microwave_2009}.

The principal aim of this study is coherent control of the muonium spin state by efficient excitation of microwave transitions. With the experiments being performed at the continuous-wave (CW) muon source at PSI, only one single muonium center at a time is present in the investigated sample. Accordingly, coherent microwave effects that translate into recorded muon decay asymmetries correspond to coherent dynamics of a single muon coupled to a single electron averaged over the ensemble of isolated implantation sites at distinct times. In this way, we aim to establish building blocks for elaborate experimental schemes to unravel complementary, \textmu{}SR-silent information on the detailed structure of the hydrogen isotopes that form upon muon implantation. The achieved performance and experimental capabilities are demonstrated with two different muonium centers, namely strongly coupled muonium formed inside SiO$_2$ and muonium with weak and anisotropic hyperfine coupling as formed in Si.

\section{\label{sec:matmet}Materials and methods}

\subsection{Experimental setup}

A dedicated probe head for microwave excitation inside the general purpose surface-muon instrument (GPS) \cite{amato_new_2017} was constructed. The probe head exposes the sample to the muon beam inside a helium flow cryostat permitting continuous operation down to 2~K. The microwave magnetic field $B_1$ at the sample is realized by a half-wave microstrip resonator \cite{johansson_stripline_1974} with resonance frequency around 4~GHz. The microwave field $B_1$ is transverse to the static external field $B_0$ that is applied along the direction of the muon beam. Further details on the technical realization of the probehead are given in Appendix \ref{app:probehead}. 

Prior to the first \textmu{}SR experiments with this resonator design at the GPS beamline, excitation of spin transitions was benchmarked extensively based on conventional ESR spectroscopy. These benchmark experiments used an open source microwave spectrometer \cite{doll_pulsed_2019} and are described in Appendix \ref{app:ESR}. Importantly, these experiments outline the potential use of our half-wave strip design for conventional ESR spectroscopy.

During \textmu{}SR experiments, microwave pulses with power up to 100 W were injected to the probehead. The pulses were triggered by a hardware event when a muon stops inside the sample, which introduces a delay due to cabling and electronics. The shortest possible delay between the arrival of the muon and the arrival of the microwave pulse in the sample was 289 ns for the experiments with SiO$_2$ and 305 ns for the experiments with Si. The microwave pulse-forming setup allows for adjustments in the pulse timing and in the microwave phase offset among multiple pulses. A detailed description of this setup is given in Appendix \ref{app:timing}.

\subsection{Samples and experimental conditions}

The SiO$_2$ quartz sample had dimensions of 10x10x0.5 mm and Z-cut orientation (MTI corporation, USA). The intrinsic Si (100) sample was previously studied and showed formation of anisotropic muonium \cite{prokscha_low-energy_2012}. This 0.4~mm thick sample had approximately square shape with 10~mm side length. Experiments on SiO$_2$ were performed at a temperature of 280 K, whereas Si was investigated at 50 K. During the measurements, the microwave pulses increased the temperature that is measured in the vicinity of the sample, especially for the Rabi experiments with long pulse durations beyond 1 \textmu{}s. To counteract microwave heating and maintain the sample at a fixed temperature, the temperature of the helium cooling gas in the flow cryostat was reduced by 20~K for SiO$_2$ and by 10~K for Si with respect to the sample temperature setpoints at 50 K and 280 K, respectively.



\subsection{Data analysis}
The measured muon decay asymmetries were analyzed by means of curve fitting via minimization of the weighted least-square error $\chi^2$. For this purpose, the iminuit python package was used, which is a python frontend to the established Minuit library \cite{james2004minuit}. All stated uncertainties in model parameters extracted in this way correspond to 68\% confidence intervals, thus representing ordinary $1\sigma$ standard deviations. A different procedure was applied for model fitting of a TF DEMUR experiment on Si. For this particular experiment, data fitting was performed using musrfit \cite{suter_musrfit_2012}, which also utilizes the Minuit library. A higher-level two-parameter fit to the model parameters extracted from musrfit was performed via an iteratively refined grid search. The resultant 2D maps of the weighted least-square error $\chi^2$ provided the best-fit parameters as well as the 68\% confidence intervals \cite{avni_energy_1976}.

Interpretation of the experimental results included numerical and analytical methods to solve the time-dependent spin evolution of the investigated electron-muon systems. Numerical calculations were performed using the SPIn DYnamics ANalysis (SPIDYAN) library \cite{pribitzer_spidyan_2016} in Matlab\textsuperscript{\textregistered}. Analytical expressions, namely all equations in Appendix \ref{app:analytical}, were computed with a dedicated library \cite{spinop} for the underlying product operator calculations \cite{sorensen_product_1984} in Mathematica\textsuperscript{\textregistered}.

\section{\label{sec:results} Results and Discussion}
\subsubsection{\label{sub:quartz} Isotropic muonium in SiO$_2$}
\begin{figure}[t!]
	\centering
	\includegraphics{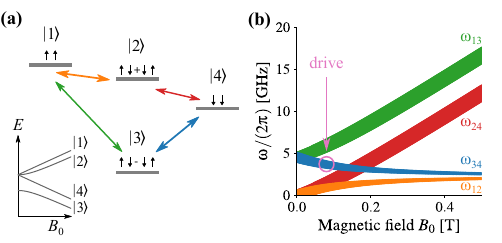}
	\caption{Effective two-level dynamics in muonium with strong electron-muon coupling of 4.5 GHz. \textbf{(a)} Energy level diagram with relevant transitions color coded. The eigenstates (not normalized) at zero applied field are indicated below the level number. The bottom left inset shows the corresponding Breit-Rabi diagram. \textbf{(b)} Transition frequencies $\omega_{ij}$ as a function of the applied magnetic field, with line thickness proportional to the Rabi frequency for a transverse excitation field. The magenta circle marks the region where the $\omega_{34}$ transition was driven.}
	\label{fig:qlevels2}
\end{figure}

Muonium formed in SiO$_2$ served as benchmark for microwave-driven two-level spin dynamics. This muonium system is well-known for it's large isotropic hyperfine coupling at room temperature that is close to the vacuum muonium value. At magnetic fields relevant in this study, anisotropic contributions are comparably small and result in small field-dependent corrections to the isotropic value \cite{brown_precision_1980}. With the dominant hyperfine interaction of 4.5 GHz, the muon and electron spins combine into singlet and triplet states. The energy levels of the singlet $\ket{3}$ and the triplets $\ket{1}$, $\ket{2}$, and $\ket{4}$ are shown in Fig. \ref{fig:qlevels2}(a), including their dependence on an applied static field $B_0$. 

An oscillating magnetic field $B_1$, transverse to $B_0$, causes phase-coherent Rabi oscillations if the oscillation frequency $\omega_{\mathrm{uw}}$ is in resonance with a transition $\omega_{ij}$. Figure \ref{fig:qlevels2}(b) shows the resonance frequencies $\omega_{ij}$ and the line thickness denotes the proportionality factor $\gamma_{ij}$ that determines the Rabi oscillation frequency $\omega_1 = \gamma_{ij}\cdot B_1$. The transition $\omega_{34}$ is well separated from other transitions in the region around 80 mT that we target in this study (magenta circle). Consequently, the spin dynamics of this four-level system can be treated as an effective two-level system. Within a small detuning $\Omega = \omega_{34} - \omega_\mathrm{uw}$ between the drive and the transition, the effective Rabi oscillation frequency increases progressively with $|\Omega|$, since $\omega_\mathrm{eff}^2 = \omega_1^2 + \Omega^2$, while the oscillation amplitude decreases as $|\Omega|$ becomes larger \cite{schweiger_principles_2001}.

Experimentally, the resonance frequency $\omega_{34}$ was varied by changing the magnetic field, while the microwave frequency was fixed. The muons were implanted with their initial polarization along the magnetic field, such that the longitudinal spin polarization of the formed muonium center can be recorded directly in the time domain. Each individual muon stopped inside the sample was followed by a 1.25 \textmu{}s long full-power microwave pulse after a delay of $t_\mathrm{p} = 491$~ns. As a result, the muon decay asymmetry $A(t)$ provided the muon polarization before, during, and after the resonant pulse. Note that not every stopped muon binds with an electron, such that the asymmetry $A(t)$ also contains a diamagnetic fraction that is small for SiO$_2$ \cite{brewer_muonium_1981}.

\begin{figure}[t!]
    \centering
    \includegraphics{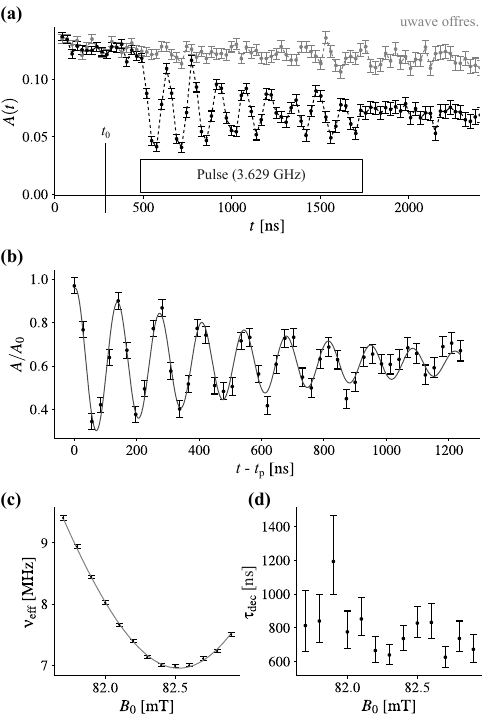}
    \caption{Rabi oscillations of the $\omega_{34}$ transition in isotropic muonium with 3.629 GHz microwaves. \textbf{(a)} Time-domain asymmetry $A(t)$ with pulse on resonance at 82.5 mT (black) and off resonance at 90 mT (gray). The dashed lines are guides to the eye and $t_0$ is the earliest possible pulse arrival time. \textbf{(b)} Time-domain asymmetry during microwave drive at 82.2 mT, normalized data (black) and fit (gray). The fit is an exponentially damped oscillation with frequency $\nu_\mathrm{eff}$ = 7.40 $\pm$ 0.03 MHz, decay time $\tau_\mathrm{dec}$ = 660 $\pm$ 80 ns, amplitude $A_\mathrm{osc}$ = 0.35 $\pm$ 0.02, and with vertical offset $A_\mathrm{static}$ = 0.613 $\pm$ 0.006. \textbf{(c,d)} Evolution of Rabi frequency $\nu_\mathrm{eff}$ and damping time $\tau_\mathrm{dec}$ with magnetic field $B_0$ (black). The gray curve in (c) is a fit of the theoretical expectation to $\nu_\mathrm{eff}$ with $\nu_1$ = 6.95 $\pm$ 0.01 MHz and $\Omega / 2\pi = \Tilde{\gamma} (B_\mathrm{res} - B_0)$, where $\Tilde{\gamma}$ = 7.67 $\pm$ 0.07 MHz/mT and $B_\mathrm{res}$ = 82.525 $\pm$ 0.005 mT. For a field of 82.2 mT, the fit reveals $\Omega / 2\pi$ = 2.496 $\pm$ 0.042 MHz.}
    \label{fig:Rabis}
\end{figure}

Exemplary raw data at small (black) and at large (gray) detuning are shown in Fig. \ref{fig:Rabis}(a) and display a pronounced effect due to the pulse. The asymmetry during the pulse is shown in Fig. \ref{fig:Rabis}(b), where the raw data were normalized by the polarization $A_0$ before the pulse. The data clearly show several coherent Rabi oscillations and follow the exponentially decaying oscillation indicated by the fit (gray). At this particular field, the effective Rabi oscillation frequency $\nu_\mathrm{eff}$ was 7.4~MHz, which means that it takes 68~ns to maximally invert the population on the $\omega_{34}$ transition. To the best of our knowledge, such fast Rabi oscillations have never been achieved with muonium around 4 GHz, neither for larger microwave cavities \cite{favart_precision_1971, nishimura_rabi-oscillation_2021} nor for a conceptually similar half-wave strip design \cite{lord_microwave_2009}.

The influence of the resonance offset $\Omega$ is exemplified in Fig.~\ref{fig:Rabis}(c), which shows the dependence of the oscillation frequency $\nu_\mathrm{eff}$ on the magnetic field $B_0$. These frequencies were determined not only accurately from the experimental data, but were also well fitted by the theoretical expectation shown as a solid gray line. The fit yields the on-resonance field $B_\mathrm{res} = 82.525$~$\pm$~0.005~mT and Rabi frequency $\nu_1 = 6.95$~$\pm$~0.01~MHz that translates into a microwave driving field amplitude $B_1 = 0.96$~mT, using the value of $\gamma_{34}$ at the magnetic field $B_\mathrm{res}$. The resonance field $B_\mathrm{res}$ can also be inferred from the amplitudes of the oscillatory part $A_\mathrm{osc}$ and of the non-oscillatory part $A_\mathrm{static}$, as illustrated in Fig. \ref{fig:supplfits} in the Appendix. However, this procedure gave less precise results compared to those obtained from the precession frequency.

To this point, our experiments are in excellent agreement with the theoretical predictions. However, all the experimental Rabi oscillations were significantly damped. Figure \ref{fig:Rabis}(d) reveals that the extracted exponential decay time $\tau_\mathrm{dec}$ is virtually independent of the magnetic field $B_0$. This excludes intrinsic spin relaxation via the decoherence time $T_2$ of the $\omega_{34}$ transition as the principal cause of the observed damping, since a pronounced dependence of the damping on the resonance offset $\Omega$ is expected for driven spins \cite{schweiger_principles_2001}. Rather, we ascribe the damping to spatial inhomogeneity in the microwave driving fields $B_1$ that is expected for the half-wave resonator used in the experiments. 

The Rabi decay, caused by the geometry of our probe, masks the intrinsic spin relaxation time of the isolated muonium center. Moreover, the extra damping limits the precision for determination of the oscillation frequency $\nu_\mathrm{eff}$. Thus, experimental techniques to mitigate this extra damping are of high interest. In principle, based on the rotary echo that is well established in NMR \cite{solomon_rotary_1959}, one could undo the dephasing due to $B_1$ inhomogeneity by inversion of the microwave phase at specific timings. Using this approach, however, the overall pulse duration and power should still be constrained to avoid excessive sample heating.

\begin{figure}
    \centering
    \includegraphics{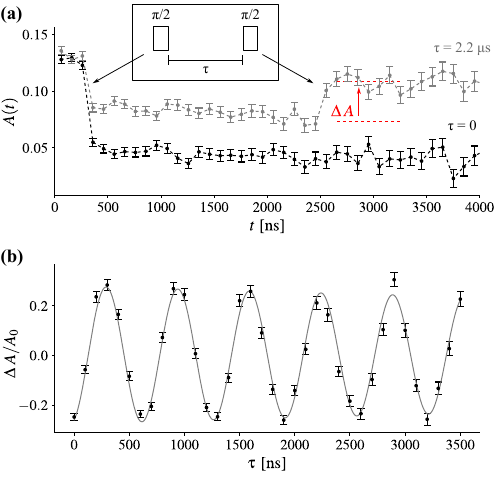}
    \caption{Ramsey fringes generated by two $\pi$/2 pulses at a delay $\tau$, using 3.805 GHz microwaves at $B_0$ = 61.14 $\pm$ 0.03 mT. \textbf{(a)} Time-domain asymmetry for $\tau$ = 0 (black) and $\tau$ = 2.2 \textmu{}s (gray). The red arrows denote the pulse timing and dashed lines are guides to the eye. \textbf{(b)} Variation $\Delta$ in normalized asymmetry $A/ A_0$ for delays $\tau$ up to 3.5 \textmu{}s (black) and fit (gray), which is an exponentially damped oscillation with frequency $\Omega / 2\pi$ = 1.540 $\pm$ 0.003 MHz, decay time $\tau_\mathrm{dec}$ = 21 \textmu{}s with asymmetric error of [-6, 14] \textmu{}s, amplitude $A_\mathrm{osc}$ = 0.278 $\pm$ 0.009, and with vertical offset of 0.002 $\pm$ 0.004. An integration window of 960 ns duration was used to extract the polarization after the second pulse. The increment in $\tau$ was 100 ns.}
    \label{fig:Ramsey}
\end{figure}


An elegant method to mitigate the problems of $B_1$ inhomogeneity and admissible microwave power is provided by pulsed techniques. With the unprecedented performance of our probe head, pulsed techniques on muonium can be realized with high efficiency. A coherent superposition state between levels ${\lvert 3 \rangle}$ and ${\lvert 4 \rangle}$ is created by a $\pi/2$ pulse that is only 30 ns long. Ramsey type of experiments \cite{ramsey_molecular_1950} that employ a pair of $\pi/2$ pulses with a delay $\tau$ between them can be easily implemented. In this experiment, the free evolution of the excited transition during the time interval $\tau$ gives rise to polarization fringes after the second pulse. These so-called Ramsey fringes are directly proportional to $\cos(\Omega \tau)$, while the damping of the fringes corresponds to the intrinsic spin relaxation time. Therefore, both the detuning $\Omega$ and the intrinsic coherence time can be accessed experimentally by variation of $\tau$.

Raw data of the Ramsey experiment are illustrated in Fig.~\ref{fig:Ramsey}(a), which shows the muon decay asymmetry $A(t)$ for delays $\tau = 0$ (black) and $\tau = 2.2$~\textmu{}s (gray). For $\tau=0$~\textmu{}s, the two $\pi/2$ pulses perform an overall rotation of $\pi$. The recorded asymmetry thus shows a step at the arrival time of the first pulse at $t_\mathrm{p} = 291$~ns. For the two pulses separated by $\tau = 2.2$~\textmu{}s (see inset), the creation of a coherent superposition state results in half the step after the first pulse. The step $\Delta A$ (red) after the second pulse depends on the acquired relative phase between the microwave and the freely precessing spins during $\tau$. For the delay here, $\Omega \tau$ was an odd multiple of $\pi$, which results in a rotation back to the initial state.

Ramsey fringes obtained from a series of measurements with different values of $\tau$ are shown in Fig.~\ref{fig:Ramsey}(b). Each data point (black) originates from a trace as in Fig.~\ref{fig:Ramsey}(a) and represents the step $\Delta A / A_0$, where $A_0$ is the asymmetry before the first pulse. The resultant Ramsey oscillation is virtually undamped, allowing the determination of $\Omega$ with a precision of 3 kHz (gray line). The less precisely determined decay time $\tau_\mathrm{dec}$ of 21 \textmu{}s retrieved upon data fitting is in reasonable agreement with the decay observed of the total longitudinal polarization of 22 \textmu{}s. We thus conclude that the coherence time $T_2$ on the $\omega_{34}$ transition is limited by the mean $T_1$ relaxation time observed on all the transitions. In this context, it is important to state that a likely limitation for $T_2$ could stem from (i) naturally abundant $^{29}$Si nuclear spins that surround the muonium center \cite{isoya_epr_1983} or (ii) hopping of the center between different sites \cite{baryshevskii_quadrupole_1983, barsov_temperature_1986}, where the latter is more likely at the studied temperature.

In order to compare to Rabi oscillations, a Ramsey data set at the same microwave frequency and magnetic field as in Fig. \ref{fig:Rabis}(b) has been acquired. Moreover, the delay $\tau$ was only incremented up to 1.2 \textmu{}s, which corresponds to the pulse duration in the Rabi experiment. The resultant 13-point Ramsey oscillation is shown in the Appendix in Fig. \ref{fig:supplramsey}. From this Ramsey experiment, the resonance offset $\Omega$ was determined as 2.472 $\pm$ 0.006 MHz. From the Rabi experiments in Fig. \ref{fig:Rabis} that also contained 13 data sets, a resonance offset $\Omega$ = 2.496 $\pm$ 0.042 MHz was obtained for the same applied field. The precision on $\Omega$ was thus enhanced by almost an order of magnitude when changing from the Rabi to the Ramsey experiment. We therefore conclude that the Ramsey experiment is better suited for precision metrology on muonium. Moreover, the long integration windows in Ramsey experiments facilitate faster acquisition times. In our experiments, a single Ramsey trace was actually acquired five times faster than a Rabi trace.

One of the drawbacks related to precise determination of $\Omega$ from Rabi oscillations is that the microwave $B_1$ field strength needs to be extracted precisely from a series of experiments in the vicinity of the resonance. In the Ramsey approach, the field strength $B_1$ only needs to be known to the point where appropriate $\pi/2$ pulses can be set up. At the expense of longer pulse durations, resilience to the resultant pulse errors could even be improved by using more advanced excitation pulses, such as composite \cite{levitt_composite_1986, clayden_spin_2012} or frequency-swept microwave pulses \cite{doll_wideband_2017}.

It is worth to distinguish the Rabi experiments performed here at non-zero magnetic field to recent zero-field Rabi experiments on muonium in noble gases \cite{nishimura_rabi-oscillation_2021}. These experiments aim for precise measurements of the vacuum muonium hyperfine splitting to test the predictions of the standard model and emergent theories. At present, it is expected that the experimental precision will allow to constrain the hadronic vacuum polarization and electro-weak contributions \cite{strasser_precision_2025}. For the experiments in zero magnetic field reported in Ref. \cite{nishimura_rabi-oscillation_2021}, the absence of Zeeman terms in the Hamiltonian results in a different oscillation pattern as observed in this study. In essence, the oscillation with $\omega_\mathrm{eff}$ is multiplied by another oscillation with $\Omega$ at zero field, such that the Rabi traces contain two oscillation frequencies (see also Fig \ref{fig:zrofieldRabi}). In this limit, it is indeed possible to directly retrieve $\omega_1$ and $\Omega$ from one single Rabi trace \cite{nishimura_rabi-oscillation_2021}. However, as soon as the electron Zeeman frequency $\omega_S$ becomes larger than $\omega_1$, a single Rabi oscillation frequency $\omega_\mathrm{eff}$ results. In the zero-field limit, one would therefore expect Rabi and Ramsey experiments to be more equivalent with respect to determination of $\Omega$.

\subsubsection{\label{sub:Si} Anisotropic muonium in Si}

Bond-centered muonium in Si is an anisotropic muonium center that is formed at cryogenic temperatures \cite{patterson_anomalous_1978}. It's centered position on a Si-Si bond has been experimentally confirmed by means of muon level crossing resonance spectroscopy \cite{kiefl_29mathrmsi_1988}, which can probe electron-mediated (super-) hyperfine interactions between the muon and $^{29}$Si nuclear spins. Since the ESR transitions are not directly accessible in an ordinary \textmu{}SR experiment, only limited information is available about the $g$-factor of the electron spin transition. An earlier DEMUR study in TF geometry targeted the ESR transitions at RF frequencies \cite{blazey_double_1986}. However, line-broadening of the ESR transition critically limited the data quality. Here we show that we can achieve an enhanced spectral resolution at the microwave frequencies that are feasible with our probe. Furthermore, we outline methods to mitigate the effect of line broadening of the ESR transitions. In part, this involves excitation of multi-quantum coherence that is enabled by the large $B_1$ fields reached with our setup.

\begin{figure}
	\centering
	\includegraphics{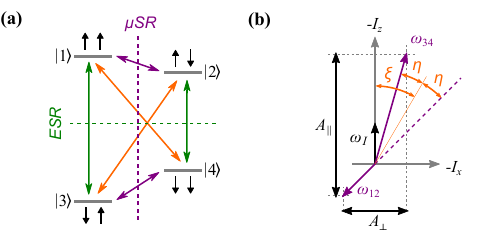}
	\caption{Four-level spin dynamics in muonium with weak anisotropic electron-muon coupling. \textbf{(a)} Energy level diagram with electron (green), muon (magenta) and electron-muon (orange) transitions. The eigenstates in the high-field limit are indicated for each of the four energy levels. \textbf{(b)} Canted muon eigenstates $\omega_{12}$ and $\omega_{34}$ (magenta) as a result of $A_\parallel$, $A_\perp$, and $\omega_I$ (black). The angles $\xi$ and $\eta$ relate the canted quantization axes to their angle bisector (orange line).}
	\label{fig:silevels2}
\end{figure}

The energy level scheme for bond-centered muonium in Fig \ref{fig:silevels2}(a) shows the ESR (green), \textmu{}SR (magenta), and multi-quantum (orange) transitions. In the studied field range, the electron spin projection $m_S = \pm 1/2$ along the applied field is a good quantum number. The muon spin state, on the contrary, emerges from the competition of its Zeeman energy $\omega_I$ with the longitudinal, $A_\parallel$, and the transverse, $A_\perp$, component of the hyperfine interaction. The transverse component $A_\perp$ causes canting of the muon spin away from the applied field, such that the muon transitions $\omega_{12}$ and $\omega_{34}$ are associated to distinct tilted quantization axes. A scheme is shown in Fig. \ref{fig:silevels2}(b), where the canting of the muon states (magenta) is parametrized by the angles $\xi$ and $\eta$ (orange) that are close to our experimental parameters. 
The canted muon states have characteristic beatings in the longitudinal and transverse muon polarization and \textmu{}SR can retrieve the parameters $\omega_I$, $A_\parallel$, and $A_\perp$. For the electron spin, resonant microwaves will not only drive Rabi oscillations on the single-quantum transitions [green in Fig \ref{fig:silevels2}(a)], but also on multi-quantum transitions (orange) with a reduced efficiency, namely weighted by $\sin(\eta)$.

\begin{figure}
    \centering
    \includegraphics{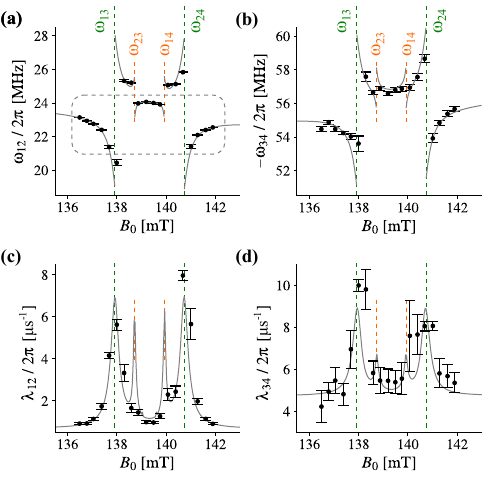}
    \caption{Transverse-field DEMUR of bond-centered muonium in Si. \textbf{(a,b)} Field dependence of muon precession frequencies $\omega_{12}$ in panel (a) and $\omega_{34}$ in panel (b) with a 3.9 GHz microwave drive. Frequency data points (black) extracted from experimental data and fit (solid gray lines) to both frequencies based on analytical solution that is parametrized by $g_e$, $w_1$, $A_\parallel$, and $A_\perp$. With $A_\parallel / 2\pi$ = 67.6 MHz and $A_\perp / 2\pi$ = 35.6 MHz determined by ordinary \textmu{}SR, the remaining free parameters are $g_e$ = 1.9999 $\pm$ 4e-4 and $2\omega_1 / \gamma_\mathrm{e}$ = 677 $\pm$ 11 \textmu{}T. Due to discontinuities in the analytical solution, 2x6 data points at specific fields were omitted, namely at the fields where $\omega_{12}$ is outside the dashed gray box in panel (a). \textbf{(c,d)} Corresponding field dependence of damping rates $\lambda_{12}$ in panel (c) and $\lambda_{34}$ in panel (d). The gray curves were extracted from numerical simulations with electron relaxation rate $1/T_2$ = 13.2 \textmu{}s$^{-1}$ and transition-specific muon relaxation rates of 0.95 and 5 \textmu{}s$^{-1}$ for $\lambda_{12}$ and $\lambda_{34}$, respectively.
    }
    \label{fig:TFDEMUR}
\end{figure}
Resonant excitation of the electron spin transitions alters the muon spin precession frequencies, which is the principle of the DEMUR experiment. Experimental results in TF geometry for the microwave-driven muon frequencies $\omega_{12}$ and $\omega_{34}$ at different magnetic fields $B_0$ along the (100) direction are shown in black in Fig. \ref{fig:TFDEMUR}(a) and (b), respectively. 
The gray curves are best fits to analytical expressions \cite{jeschke_generation_1996,schweiger_principles_2001} of the relevant spin Hamiltonian neglecting relaxation effects (see Appendix \ref{app:analytical}). The fitting parameters were the electron $g$-factor that determines the horizontal position of the characteristic spectral pattern (field dependence), and the microwave $B_1$ field that determines the vertical \emph{Rabi splitting} in muon frequencies. The other parameters of the spin Hamiltonian were determined independently: The hyperfine parameters $A_\parallel$ and $A_\perp$ were fixed to the values obtained from a muon precession measurement without microwaves (see also Appendix \ref{app:DEMUR}), while $\omega_I$ was determined by the internal reference from muons that stop in aluminium parts of the microwave resonator and from the diamagnetic fraction in Si. Note that certain data points had to be excluded from the fit (see caption and dashed gray box), since the proximity to discontinuities in the analytical model inhibited a smooth minimum of the $\chi^2$ fitting target. 

The electron $g$-factor $g_\mathrm{e}$ = 1.9999 $\pm$ 4e-4 retrieved in this way agrees with a previously reported value of 2.0 $\pm$ 2e-2 in Ref.~\cite{blazey_double_1986} and has a precision that exceeds the previous value by more than an order of magnitude. At this level of detail, it can even be compared to the ESR $g$-factor of 1.9992 for bond-centered hydrogen in Si with magnetic field along the (100) direction \cite{gorelkinskii_electron_1991}. The electron properties of bond-centered hydrogen and muonium in Si are virtually identical \cite{van_de_walle_structural_1990,porter_muonium_1999}, such that we relate the small deviation between the DEMUR and ESR $g$-factors to overall experimental uncertainty. 

The detailed line shape of the DEMUR spectrum entails four distinct microwave resonances that are labeled accordingly, namely the single-quantum electron spin transitions $\omega_{13}$ and $\omega_{24}$ (dashed green) and the multi-quantum electron-muon transitions $\omega_{23}$ and $\omega_{14}$ (dashed orange). Owing to their larger transition moment, the single-quantum transitions are, to a large part, responsible for the observed pattern. The multi-quantum transitions result in additional features that are most pronounced in $\omega_{12}$ and are well captured by the experimental frequencies.

Ideally, one would expect that the muon frequencies each split into two lines when driven resonantly by microwaves \cite{estle_theory_1983}. For our sample, however, such Rabi splittings were not observed due to dephasing. The corresponding data for the muon damping rates $\lambda_{12}$ and $\lambda_{34}$ are shown in Fig. \ref{fig:TFDEMUR}(c) and (d), respectively. Especially in the vicinity of the single-quantum microwave transitions, the muon frequencies $\omega_{12}$ and $\omega_{34}$ become strongly damped. The cause for this resonant decay of the muon spins is the line broadening of the ESR transitions. For reference, the gray curves show the corresponding damping that was extracted from numerical simulations of the spin system with $1/T_2$ = 13.2 \textmu{}s$^{-1}$ and transition-specific muon relaxation rates of 0.95 and 5 \textmu{}s$^{-1}$ for $\lambda_{12}$ and $\lambda_{34}$, respectively. The fast dephasing of the electron transitions results in accelerated dephasing of the muon transitions. In such a situation, Rabi splittings are difficult to observe and the less damped muon frequencies in the vicinity of the microwave resonance contain the relevant information. During the experiment, one therefore needs to make an adequate choice of the driving field $B_1$ with respect to the scanned magnetic field range. In the TF DEMUR experiments reported here, this was accomplished by deliberate reduction of the incident microwave power to 6 W.

Key candidate for the fast dephasing of the ESR transitions is the bath of spectroscopically unresolved $^{29}$Si nuclear spins surrounding the muonium center. It is well established that a bath of (nuclear) spins interacting with a localized central (electron) spin induces dephasing via spectral diffusion \cite{herzog_transient_1956}. In contemporary literature, this problem is often referred as the central spin dephasing problem and is of critical importance for the success of envisioned solid-state quantum computation schemes \cite{de_sousa_theory_2003, witzel_quantum_2012}. Recently, this problem has also been addressed in the context of coupled muon-fluorine systems \cite{wilkinson_information_2020}. 

In our experimental setting, the localized muonium center is embedded into a matrix of naturally abundant $^{29}$Si nuclear spins, with the spatially close nuclear spins resulting in (static) inhomogeneous line broadening and the further distant nuclear spins considered as the bath spins. Notably, there is only one single muonium center present in the sample, such that dipolar couplings to other muonium centers do not need to be considered. In addition to $^{29}$Si nuclear spins, there could be other bath spins related to native paramagnetic defect centers in our intrinsic Si sample. As an example, $T_1$ relaxation times below 1 \textmu{}s are expected for shallow donors at the investigated temperature of 50 K \cite{castner_direct_1962}. Formation of muonium in proximity of a shallow donor could therefore result in sub-microsecond dephasing times of the muonium-bound electron, or even in electron capture \cite{scheuermann_radio-frequency_1997}. In principle, even the spur of the arriving muon itself could be a source of stable paramagnetic centers at very low temperatures \cite{bucci_low_1981}. Detailed investigations of the dephasing of the muonium-bound electron are therefore considered highly valuable.

For more insight into dephasing, experiments with full-power microwave pulses in LF geometry were performed. 
In principle, one would expect the same Rabi splittings in LF mode as observed in TF mode due to the canted muon states (see also Appendix \ref{app:DEMUR}). These Rabi splittings were, however, not of primary interest in LF mode due to the fast dephasing mentioned above. Rather, we aimed for Rabi oscillations in the muon polarization by a strong drive on one of the multi-quantum transitions and extensions thereof using Ramsey-type of experiments. 
Further details related to setting up these experiments are given in Appendix \ref{app:DQ}.




\begin{figure}
    \centering
    \includegraphics{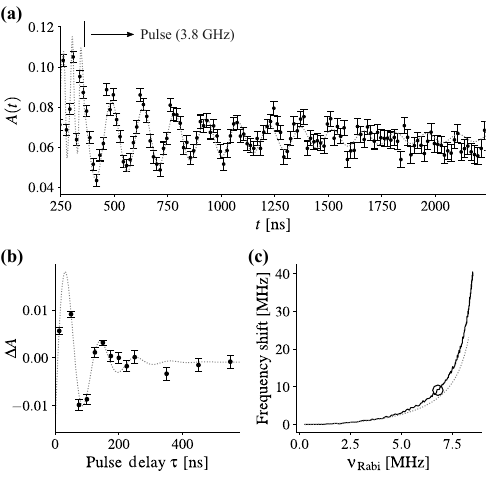}
    
    
    \caption{Drive of double-quantum transition $\omega_{14}$ in longitudinal field geometry with 3.9 GHz microwaves. \textbf{(a)} Rabi oscillations at the experimentally optimized field of 140.2 mT (see also Fig. \ref{fig:LFDQ_raw}), showing experimental data binned for visualization purposes (black) and fit before and during the pulse (dashed gray). Before the pulse, the fit incorporates the muon frequencies $\omega_{12}$ and $\omega_{34}$. During the pulse, the fit is a (vertically offset) damped oscillation with oscillation frequency $\nu_\mathrm{eff}$ = 6.79 $\pm$ 0.02 MHz and exponential decay constant $1/\tau_\mathrm{dec}$ = 1.38 $\pm$ 0.14 \textmu{}s$^{-1}$. The pulse arrival time indicated by the black horizontal line at 359.72 ns minimizes the discontinuity where the two fits coincide. \textbf{(b)} Ramsey fringes as a function of pulse delay $\tau$ at 140.2 mT. The pulse lenghts were 72 ns and the second pulse's phase was offset by either 0 or 180 degrees (see also Fig. \ref{fig:LFDQ_pcyc}). The fit (dashed gray) to experimental data (black) is a damped oscillation with oscillation frequency $\Omega$ = 9.1 $\pm$ 0.2 MHz and exponential decay constant $1/\tau_\mathrm{dec}$ = 13.2 $\pm$ 1.2 \textmu{}s$^{-1}$. \textbf{(c)} Numerical simulation of the drive-induced frequency shift of the $\omega_{14}$ transition with respect to it's static value. The abscissa shows the on-resonance Rabi oscillation frequency in $<\hat{I}_z>$. Data extracted from numerical simulation with $g_e$ = 1.9999, $A / 2\pi$ = 67.6 MHz, $B / 2\pi$ = 35.6 MHz, with step-wise progression of field-strength $B_1$ and sweep of $B_0$ around the resonance. The black circle shows the frequency shift of 9.11 MHz with a Rabi oscillation frequency $\nu_1$ of 6.79 MHz that is obtained for a driving field $B_1$ = 2.735 mT. The dashed gray line is the frequency shift of $\omega_{14}$ obtained analytically based on the frame tilting angle $\chi$ (see Appendix  \ref{app:analytical}). As expected, the analytical model is accurate for small $\nu_\mathrm{Rabi}$ and looses applicability as $\nu_\mathrm{Rabi}$ increases.
    }
    \label{fig:LFdrive14}
\end{figure}


Rabi oscillations in raw data (black) by driving the double quantum transition $\omega_{14}$ at a field of 140.2 mT are shown in Fig. \ref{fig:LFdrive14}(a). The dashed gray curve is a fit to the data before and during the pulse. Before the pulse, the fit contains the muon frequencies $\omega_{12}$ and $\omega_{34}$, with the former being the dominant component. During the pulse, the fit is a vertically-offset damped oscillation with frequency $\nu_\mathrm{eff}$ = 6.79 $\pm$ 0.02 MHz and damping $1/\tau_\mathrm{dec}$ = 1.38 $\pm$ 0.14 \textmu{}s$^{-1}$. Interestingly, the damping as well as the oscillation frequency are both rather comparable to the results in SiO$_2$ in Fig. \ref{fig:Rabis}. For the similarity in oscillation frequencies, the transition moments are related roughly via $\gamma_{14}^{\; \mathrm{Si}} \approx \gamma_{34}^{\;\mathrm{SiO}_2} / 2 \approx \gamma_\mathrm{e} / 4$, where $\gamma_\mathrm{e}$ is the gyromagnetic ratio of the electron. Therefore, a much larger full power $B_1$ has been achieved for Si at 50 K than for SiO$_2$ at 280 K. Given the temperature dependence of the resonator quality factor $Q$, the better performance at a lower temperature is not surprising. From the similarity in Rabi dampings of the two different samples, we conclude that the Rabi decay was also caused primarily by spatial inhomogeneity of the driving field $B_1$. Consequently, the driven double-quantum transition had a decoherence time that is longer than the experimental Rabi decay time and the dephasing timescale of 100 ns in the TF DEMUR data.

The result of a Ramsey experiment at the same field and microwave frequency as the Rabi experiment is shown in Fig. \ref{fig:LFdrive14}(b). The experimental data points (black) originate from two data sets with an alteration of the phase of the second pulse by 180$^\circ$. In this way, the step change $\Delta A$ caused by the two-pulse block could be obtained directly (see Appendix \ref{app:DQ}). The dashed gray line is an exponentially damped oscillation with frequency $\Omega$ = 9.1 $\pm$ 0.2 MHz and damping constant $1/\tau_\mathrm{dec}$ = 13.2 $\pm$ 1.2 \textmu{}s$^{-1}$. Since the Ramsey fringes decay much faster than the Rabi oscillations, the situation is opposite to the experiments in SiO$_2$. Moreover, the detuning $\Omega$ at the probed field was substantial and corresponds to a field offset on the order of $\pm$0.3 mT.

In order to explain the field offset, the $B_0$ and $B_1$ field dependence of Rabi oscillations has been simulated numerically. From the resultant $\nu_\mathrm{eff}(B_0, B_1)$ maps, the resonance offset $\Omega$, between the driving frequency and $\omega_{14}$, that minimizes $\nu_\mathrm{eff}$ for each $B_1$ value has been extracted. While for a two-level system, the minimum $\nu_\mathrm{eff}$ always occurs at $\Omega$ = 0, this was not the case for the simulated four-level system. The resultant dependence of the resonance offset with respect to the corresponding effective \emph{on-resonance} Rabi frequency $\nu_\mathrm{Rabi}$ is shown by the black curve in Fig. \ref{fig:LFdrive14}(c), while the dashed gray curve is an analytical approximation (see Appendix \ref{app:analytical}). As readily seen, there is a shift of the effective resonance frequency that increases with the driving field strength $B_1$. The data point marked with a circle indicates a frequency shift of 9.11 MHz that is obtained at the experimental Rabi oscillation frequency of 6.79 MHz. This frequency shift is due to off-resonant excitation of transitions that share an energy level with the $\omega_{14}$ transition. In our setting, the most important off-resonant contribution originates from the adjacent single-quantum ESR transition $\omega_{24}$. 

In general, one would expect the combination of fast dephasing on ESR transitions and comparably narrow \textmu{}SR lines to be quite common for muonium with weak electron-muon hyperfine interaction. First, the muon's gyromagnetic ratio $\gamma_\mu$ is much smaller than the coupled electron's $\gamma_\mathrm{e}$, such that the electron spin is more prone to dephasing via dipolar couplings to surrounding spins. Second, the electron wavefunction has generally a larger spatial extent than the muon wavefunction, which is another potential source of line broadening from other spins that coordinate within the electron cloud.

To better understand the source of dephasing in our experiments, we focus on the absence of fast damping in the resonantly driven LF Rabi oscillations. The LF Rabi oscillations had roughly four times the driving amplitude of the TF DEMUR experiments. With this strong drive, the effective Rabi oscillation frequency on the $\omega_{14}$ transition of 6.79 MHz was comparable to the transition's line width of $\delta = 1/(\pi \tau_\mathrm{dec})$ = 4.2 MHz. The concurrently excited single-quantum transition $\omega_{24}$ was even driven at a rate that significantly surpassed the line width. As discussed in the following, coupling of the electron spin to the surrounding matrix spins becomes effectively decoupled under such driving conditions. 

For muonium embedded in a matrix of $^{29}$Si spins at natural abundance, there are both homogeneous and inhomogeneous contributions to ESR line broadening. Inhomogeneous line broadening is caused by $^{29}$Si spins in close spatial proximity to the muonium center, where the nuclear resonance frequencies experience distinct shifts due to the nearby electron spin. As exemplified in Appendix \ref{app:narrowing} for the investigated four level system, inhomogeneous contributions are narrowed under a strong drive at the double-quantum transition. The underlying narrowing mechanism is present in any Rabi experiment where the derivative of the Rabi frequency $\nu_\mathrm{eff}$ with respect to the resonance offset $\Omega$ vanishes on resonance [see for instance Fig. \ref{fig:Rabis}(b)]. Homogeneous line broadening is caused by spin diffusion to further distant bath spins with near-degenerate energy levels, e.g spins beyond the nuclear spin diffusion barrier \cite{blumberg_nuclear_1960, khutsishvili1962spin}. This dynamic contribution to electron spin relaxation becomes decoupled by a continuous drive with strength beyond the relaxation rate (dynamical decoupling) \cite{hanson_coherent_2008, dobrovitski_decay_2009}.

In summary, the double-quantum drive results in considerable ESR line narrowing. Spectroscopic limitations related to the ESR line width can thus be alleviated for the solid-state spin system at hand. In such a situation, Rabi oscillations can be used to precisely determine the position of the driven double-quantum transition. Since the strong drive itself might shift the involved transition frequencies considerably, it is advantageous to record the full field-dependent Rabi oscillation spectrum, e.g. $\nu_\mathrm{eff}(B_0)$, which includes the equivalent dynamics on the zero-quantum transition. Accordingly, the $g$-factor as well as further spectral details of the electron spin could be determined via the symmetry of the resultant multi-quantum Rabi spectrum.

\section{Summary and Outlook} \label{sec:summ}

Overall, our study presents unprecedented performance for microwave excitation of muonium centers. Our results clearly demonstrate that it is possible to coherently control the electron-muon spin states. The Ramsey experiments presented in this study constitute the most basic variant of such techniques. Moreover, we expect that possible extensions to more elaborate pulse sequences are certainly within reach.

By contrasting experiments on effective two-level systems encountered in SiO$_2$ against four level dynamics observed in Si, we have outlined an important aspect of these experiments, namely the effect of the line width of the (hidden) microwave spectrum. For spectroscopy on effective two-level systems with strong electron-muon coupling, we anticipate that line broadening of the targeted electron-muon transition can, in general, be narrowed. First of all, microwave experiments on such strongly coupled muonium have so far been performed in the limit of narrow intrinsic line widths, which includes precision spectroscopy on vacuum muonium to test predictions of the standard model. When driving Rabi oscillations, these narrow lines become broadened by inhomogeneities of the resonant driving field. We have confirmed experimentally that this inhomogeneous broadening pathway is suppressed in a Ramsey experiment and therefore offer a less perturbing approach. 

Second, if such experiments were to be performed on systems with broader line widths, a certain suppression of homogeneous and inhomogeneous dephasing pathways is achieved by driving Rabi oscillations at large drive amplitudes. We have addressed the underlying decoupling mechanisms in the context of the sustained Rabi oscillations observed in Si on the electron-muon double-quantum transition, which was strongly damped in the absence of the concurrent microwave drive.

The situation is more intricate for muonium centers with electron-muon couplings that are of the same order as the applied microwave field. In such a case, the drive induces concurrent multi-level dynamics. The DEMUR experiment, which is based on Rabi splittings caused by hidden ESR transitions driven into resonance, is currently the principal technique dedicated to such systems. Our high-resolution DEMUR experiment on Si evidenced that ESR line broadening imposes important limitations. A critical balance between the actual effect on muon precession frequencies and the muon damping imposed by the driven electron has been obtained by detuning the microwaves off resonance. This in turn allowed for extraction of the $g$-factor with unprecedented precision. Given the susceptibility of DEMUR to the ESR line width, the development of alternate approaches is of high relevance. The experimentally demonstrated narrowing of the driven double-quantum transition up to instrumental inhomogeneities is a first step in this direction.

Besides the well-oriented muonium centers encountered in crystalline solids, we envision that coherent microwave control schemes also bear a potential with respect to related scientific problems. A first candidate are muonium radicals that are formed in (disordered) soft matter \cite{mckenzie_positive_2013}, where coherent microwave schemes could uncover detailed information content that is tied to the hidden electron. The added challenge for these disordered systems lies in distributed electron-muon hyperfine couplings of the resultant muonium spin labels, eventual anisotropies in $g$-factors, and delayed formation kinetics. Experimental schemes borrowed from ESR spectroscopy that are dedicated to hyperfine and Zeeman disorder could prove pivotal to probe the hidden electron in an efficient way. 

A second candidate are native microwave excitations intrinsic to the studied sample, such as ESR in paramagnetic systems or magnons in correlated systems. \textmu{}SR combined with a microwave stimulus could provide a complementary internal probe of such excitations. We expect such microwave pump, muon probe experiments to be particularly unique in combination with low energy muon beams that facilitate depth-resolved studies on technologically relevant films and interfaces.

\begin{acknowledgments}
This work is based on experiments performed at the Swiss Muon Source S\textmu{}S at the Paul Scherrer Institute. The research was funded from the Paul Scherrer Institute's cross initiative (project grant “Toward Quantum Technologies with X-rays and Muons”).

M. Elender and H.-P. Weber are acknowledged for assistance with the probehead design. R. Scheuermann and H. Luetkens are acknowledged for fruitful discussions on the timing and pulse forming setup.
\end{acknowledgments}

\section*{Data Availability Statement}

The code and data generated in this study have been deposited in the Zenodo database under \href{https://doi.org/10.5281/zenodo.15090279}{https://doi.org/10.5281/zenodo.15090279}.

\appendix

\clearpage 

\section{Microwave probehead} \label{app:probehead}

\begin{figure}
	\centering
	\includegraphics{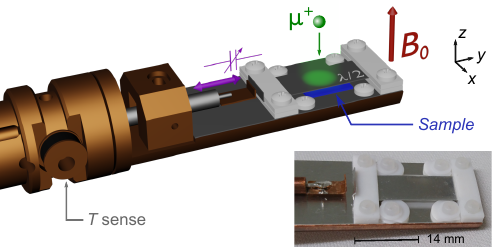}
	\caption{Microwave probe head for \textmu{}SR experiments, showing the sample (blue) underneath the half-wave strip (labelled $\lambda/2$) exposed to the muon beam (green), and with an out-of-plane magnetic field $B_0$ (red). The sample temperature is measured by the sensor on the left (Lake shore cryotronics, DT-470-CU-11A). The photograph at the bottom right shows the fabricated structure with the SiO$_2$ sample fixed by PTFE screws and washers.}
	\label{fig:probehead}
\end{figure}

This Appendix provides technical details on the employed microwave probehead. A schematic view of the half-wave resonator is shown in Fig. \ref{fig:probehead}, including a photograph of the actual setup. The half-wave strip is realized by a 50 \textmu{}m thick aluminium sheet that is 28~mm long and 8~mm wide (see $\lambda/2$ label). To form a microstrip waveguide, polytetrafluoroethylene (PTFE) spacers suspend this sheet at a distance of 1~mm above an aluminium ground plane. The sample (blue) is placed between the strip and the ground plane, where the microwave field $B_1$ along the $x$-axis is maximized \cite{shrestha_nonresonant_2019}. The muon beam (green) with full-width at half-maximum (FWHM) of 5.8~mm therefore passes through the half-wave strip before being stopped inside the sample. To minimize the number of muons stopping in the strip, we used a 50 \textmu{}m thin aluminium sheet for the SiO$_2$ sample. For the experiments with Si, a thicker sheet (300~\textmu{}m) was used to provide an in situ reference for the applied magnetic field using the signal from muons stopping in the strip. 

The size of the muon beam with respect to the length of the resonator is a critical factor, since the microwave magnetic field along the $y$-axis follows roughly a $\sin(y / (\lambda / 2)\cdot \pi )$ dependence. Within the FWHM of the muon beam, a spatial variation on the order of 5\% is expected, which we consider as a lower limit given the largely simplified analysis. The experimental Rabi oscillations for Si and SiO$_2$ were fitted by damped oscillations, whose spectral FWHM on the order of 7\% agree with our simple estimate.

The coupling between the microwave resonator and the feedline can be adjusted mechanically inside the cryostat, as conceptually illustrated in magenta. For the experiments performed at either 280~K on SiO$_2$ or at 50~K on Si, the resonator quality factor was set to 60. With the principal frequency around 4~GHz, this corresponded to resonator bandwidths on the order of 65~MHz. 

\section{ESR benchmarking of microwave resonator} \label{app:ESR}
\begin{figure}[!b]
	\centering
	\includegraphics{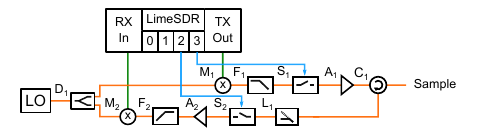}
	\caption{Schematic of the spectrometer used for pulse ESR experiments, showing microwave circuit with the labeled components listed in table \ref{tab:instr2}. The orange lines carry signals at the targeted microwave frequency. The LimeSDR software-defined radio provides control signals for switches (blue lines) and interfaces to frequency-translated signals at lower frequencies around 1-2 GHz (green lines).}
	\label{fig:SDRsetup}
\end{figure}

Pulsed ESR was used to benchmark the microwave resonator prior to \textmu{}SR experiments. The sample was the Koelsch radical with formula 1,3-bisdiphenylene-2-phenylallyl (BDPA). A mm-sized grain of this radical was attached at the sample position inside the microwave resonator. To excite the spins and record their response, an open source spectrometer based on the LimeSDR software-defined radio platform was used \cite{doll_pulsed_2019}, which can output freqencies up to 3.8 GHz. In order to access higher frequencies, its input and output signals were frequency-translated with a 5 GHz local oscillator (LO). A schematic of this setup is shown in Fig. \ref{fig:SDRsetup}, which includes the chain of devices that are listed in Table \ref{tab:instr2}.

The probehead was inserted into the cryostat of the \textmu{}SR spectrometer and held at 250 K. The field was set to 134.95 mT, where the BDPA sample was in resonance with the utilized microwave frequency of 3.7795 GHz. Spin echoes were formed by a pair of 98 ns long pulses with pulse amplitudes calibrated to perform the required $\pi$/2 and $\pi$ rotations. The gap between these two pulses was 130 ns. Each spin echo was averaged 8000 times at a repetition rate of 12 kHz. A two-step (and four-step) phase cycle on the $\pi$/2 (and $\pi$) pulse was used to select the spin echo.

\begin{table}[!b]
	\centering
	\begin{tabular}{|l|l|}
		\hline 
		Label & Device\\
		\hline \vspace{-0.4cm}
		\\\hline
		A$_1$ & Minicircuits, HPA-100W-63+\\
		A$_2$ & Minicircuits, ZX60-83LN-S+\\
		D$_1$ & Minicircuits, ZX10-2-622-S+\\
		S$_1$, S$_2$ & Minicircuits, ZASWA2-50DR-FA+\\
		L$_1$ & Minicircuits, VLM-83-2W-S+\\
		C$_1$ & JQL corp., JCC2000T4000S3R\\
		M$_1$, M$_2$ & Minicircuits, ZX05-14LH-S+\\
		F$_1$, F$_2$ & Minicircuits, VLF-5000+\\
		LO & Era Instruments, EraSynth Micro at 5 GHz\\
		\hline
	\end{tabular}
	\caption{List of components labeled in Figure \ref{fig:SDRsetup}. \label{tab:instr2}}
\end{table}

\begin{figure}[!t]
	\centering
	\includegraphics{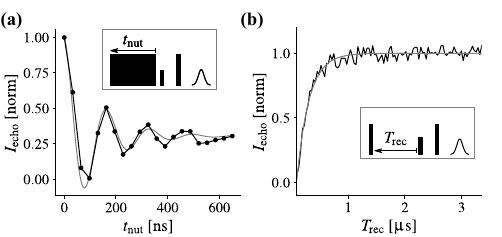}
	\caption{Pulse ESR benchmark experiments with BDPA, where the insets denote the pulse sequence. \textbf{(a)} Transient nutation experiment showing Rabi oscillations as a function of $t_\mathrm{nut}$ (black) and a fit oscillating at 6.0 $\pm$ 0.1 MHz with exponential time constant of 126 $\pm$ 16 ns and non-oscillatory offset of 0.3 $\pm$ 0.02. \textbf{(b)} Inversion recovery experiment showing experimental data (black) and exponential fit (gray) with decay time $T_1$ = 216 $\pm$ 9 ns.}
	\label{fig:ESRbench}
\end{figure}

Results from two pulse sequences are shown in Fig. \ref{fig:ESRbench}. Panel (a) shows Rabi oscillations (black) as a function of the pulse duration $t_\mathrm{nut}$, which were recorded with the transient nutation pulse sequence shown in the inset. The data were fit using a damped oscillation (gray) with Rabi oscillation frequency of 6 $\pm$ 0.1 MHz and damping time constant of 126 $\pm$ 16 ns. 
The damping of the Rabi oscillations of BDPA is due to the relaxation times that are on a comparable timescale. The longitudinal relaxation time $T_1$ was determined with the inversion recovery sequence in Fig. \ref{fig:ESRbench}(b). The experimental polarization decay (black) followed an exponential decay with $T_1 = 216 \pm 9$ ns (gray).

The oscillation frequency of BDPA is comparable to the 7 MHz Rabi frequency observed for muonium in SiO$_2$. Such benchmark experiments are thus useful for monitoring of the hardware performance offline without using the moun beam. Importantly, there is also no need for a cryostat when working with BDPA. During the design phase of the microwave probehead, we have performed such experiments at ambient temperature on a test setup with an electromagnet.

\section{Microwave timing setup} \label{app:timing}

\begin{figure}[t!]
	\centering
	\includegraphics{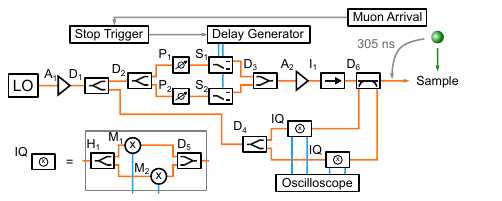}
	\caption{Microwave setup for \textmu{}SR experiments, showing microwave circuit with the labeled components listed in table \ref{tab:instr}. The orange lines carry microwave signals, while the blue lines carry pulse envelopes. See text for further descriptions. A 10~dB attenuator connected at the forward power monitor of D$_6$ has been omitted in the drawing for simplicity. }
	\label{fig:timingsetup}
\end{figure}

A microwave timing setup provided microwave pulses with well-defined timing with respect to single-muon arrival. The setup consists of (i) a delay structure that generates trigger pulses at user-defined delays, and (ii) a microwave circuit that converts these trigger pulses into microwave pulses with user-defined phases. A schematic view of this setup is given in Fig. \ref{fig:timingsetup}. The delay structure is essentially the digital delay generator shown at the top of the schematic, while the microwave circuit is depicted below, with the list of components in Table \ref{tab:instr}.

The delay structure relies on a commercial digital delay generator (Standford research systems, DG645 with option 03). This delay generator provides four trigger pulses with variable delay and duration upon an externally provided reference trigger, such that our setup allows for a maximum of four microwave pulses. The delay generator features a transition time of 85~ns at a very low timing jitter $<$30~ps with respect to the external reference trigger. For delayed trigger pulses with respect to a \emph{single muon event}, we used the standard incoming muon counter signal of GPS as a trigger. The actual definition of this muon event depends on detector veto/anti-coincidence rules and is detailed elsewhere \cite{amato_new_2017}. In brief, for the large samples used in this study, the stop trigger corresponded to arrival of a muon onto the sample (operation in \emph{no-veto} mode). 

\begin{table}[t!]
	\centering
	\begin{tabular}{|l|l|}
		\hline 
		Label & Device\\
		\hline \vspace{-0.4cm}
		\\\hline
		A$_1$ & Minicircuits, ZX60-83LN-S+\\
		A$_2$ & Minicircuits, HPA-100W-63+\\
		D$_1$ - D$_5$ & Minicircuits, ZX10-2-622\\
		D$_6$ & Minicircuits, ZGBDC35-93HP+\\
		P$_1$, P$_2$ & Analog Devices, EVAL-HMC649LP6 \\ & controlled by a Raspberry Pi\\
		S$_1$, S$_2$ & Minicircuits, ZASWA2-50DR-FA+\\
		I$_1$ & Uiy Inc, BCC3030A2T6NF, port \#3 \\ & of circulator terminated by\\ & Minicircuits, TERM-50W-183S+\\
		H$_1$ & Minicircuits, QCS-592 TB\\
		M$_1$, M$_2$ & Minicircuits, ZX05-83-S+\\
		LO & Era Instruments, EraSynth+ \\ & controlled by a Raspberry Pi\\
		\hline
	\end{tabular}
	\caption{List of components labeled in Figure \ref{fig:timingsetup}. \label{tab:instr}}
\end{table}

In order to form microwave pulses, the delay generator controls two microwave switches S$_1$ and S$_2$ that constitute different microwave phase channels. These two switches are driven by logical combinations of the four trigger pulses that are provided by the delay generator. While switch S$_1$ receives a single trigger pulse, a logical OR combination of the three other trigger pulses are directed to S$_2$. Accordingly, S$_1$ can generate a single microwave pulse of variable delay and duration, whereas S$_2$ can generate up to three such pulses. The relative microwave phase of pulses that are formed by either of these switches can be adjusted by digital microwave phase shifters P$_1$ and P$_2$ with phase resolution of 5.625$^\circ$. In this way, basic phase cycling schemes that are well established in magnetic resonance for coherence transfer pathway selection among multiple pulses are feasible. 

The microwave circuit also incorporates a number of other elements that are briefly explained in the following. The full microwave chain starts at the CW microwave source with label LO (local oscillator). The continuous microwave signal of this source passes an additional driving amplifier (A$_1$) and is split into the main excitation branch (top pathway) and a monitor branch (bottom pathway). The main excitation branch incorporates the two phase channels in-between the signal dividers D$_2$ and D$_3$. The combined signal is routed to a 100~W microwave amplifier, whose output is protected by the isolator I$_1$. Prior to injection into the probe head, the directional coupler D$_6$ provides attenuated monitor signals of the injected and reflected pulses. These monitor signals are converted down from microwave to zero frequency in the monitor arm. The frequency conversion considers both the in-phase and the quadrature component, labeled IQ in the diagram. Accordingly, the monitor branch provides information on pulse timing and phases for direct inspection with an oscilloscope.

Due to transition delays of devices and cables, there is a minimum time gap, $t_\mathrm{p}$, between the arrival of a muon and a microwave pulse at zero programmed delay. The delay between the stop trigger and the coupling structure of the resonator was determined as 183~ns using delay histograms on a fast oscilloscope. Approximately half of this delay is attributed to the delay generator, whereas the other half comes from the overall transition delay through the setup. Including an additional delay of 122~ns related to generation of the muon stop signal, a total delay of $t_\mathrm{p}=305$~ns between muon and microwave pulse arrival results. Note that due to the finite bandwidth of the microwave resonator, an additional transition time on the order of 5~ns is required for the pulse to build up inside the resonator.

All relevant parameters of the microwave setup can be controlled remotely and are integrated into the MIDAS data acquisition system \cite{MIDAS_link} operated at the Swiss muon source. The control of the microwave source as well as the microwave phase switches is relayed via a Raspberry Pi 4 computer, while the delay generator is controlled by a dedicated MIDAS device driver. 

The setup above describes the state as used in the experiments with Si. For the experiments with SiO$_2$, an earlier version of the setup was used. The microwave circuit consisted of a slightly different microwave source (Era Instruments, EraSynth micro), followed by $S_1$ and $A_2$, meaning that phase cycling and monitoring capabilities were not yet available. Moreover, a two-channel digital delay generator (Standford research systems, DG535) was used. With this setup, a slightly shorter overall delay $t_\mathrm{p} = 289$~ns was attained.

\section{Supplementary figures for SiO$_2$ experiments} \label{app:quartz}

\begin{figure}[!t]
    \centering
    \includegraphics{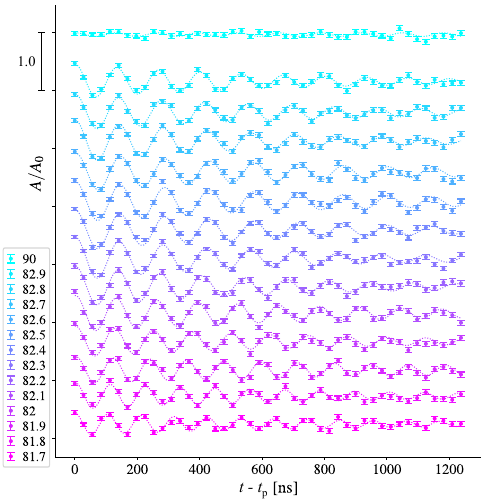}
    \caption{Rabi oscillations during microwave drive recorded at different magnetic fields (see legend for field in mT), showing binned experimental data (bars) and fits (dashed lines). The time $t_\mathrm{p}$ corresponds to the arrival of the microwave pulse.}
    \label{fig:allrabis}
\end{figure}

\begin{figure}[!b]
    \centering
    \includegraphics{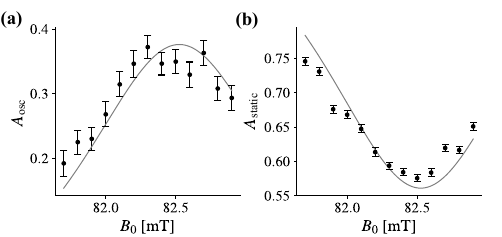}
    \caption{Magnetic field dependence of \textbf{(a)} the Rabi oscillation amplitude $A_\mathrm{osc}$ and \textbf{(b)} the constant background $A_\mathrm{static}$. The gray curves are predictions based on Eq.~\eqref{eq:2lev_amp} with best-fit polarizations $p_{34}$ = 0.38 and $p_{\Sigma}$ = 0.94 and the fitted dependencies of $\Omega$ and $\omega_\mathrm{eff}$ from Fig. \ref{fig:Rabis}(b). } 
    \label{fig:supplfits}
\end{figure}

This Appendix contains supplementary figures related to experiments on SiO$_2$, which are all discussed in the main text. The raw data and corresponding fits of all the field-dependent Rabi oscillations are shown in Fig. \ref{fig:allrabis}, where data at different magnetic fields are vertically offset according to the legend. The amplitudes of the oscillatory and non-oscillatory part of these Rabi oscillations, $A_\mathrm{osc}$ and $A_\mathrm{static}$, respectively, are shown in Fig. \ref{fig:supplfits}. The data were fit using
\begin{align}
	A_\mathrm{osc} &= p_{34} \cdot (1 - (\Omega/\omega_\mathrm{eff})^2) \label{eq:2lev_amp}
\end{align}
where $p_{34}$ is the partial polarization of the $\omega_{34}$ transition. The non-oscillatory part $A_\mathrm{static}$ of the muon polarization is determined by $A_\mathrm{static} = p_{\Sigma} - A_\mathrm{osc}$, where $p_{\Sigma}$ is the sum of all partial polarizations that are static in the absence of a resonant drive. From an experimental point of view, this reduction in $A_\mathrm{static}$ by resonant microwaves is directly accessible by time-integral acquisition techniques, as employed in earlier precision spectroscopy experiments on muonium in noble gases \cite{thompson_muonium_1973, casperson_new_1975, liu_high_1999}.

\begin{figure}[!h]
    \centering
    \includegraphics{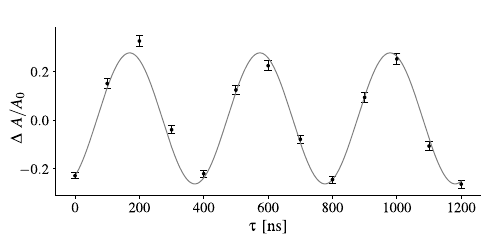}
    \caption{Ramsey fringes at 82.2 mT at experimental conditions identical to Rabi oscillations in Fig. \ref{fig:Rabis}, showing experimental data (black) and a fit (gray). The fit is an oscillation with frequency $\Omega / 2\pi$ = 2.472 $\pm$ 0.006 MHz, amplitude $A_\mathrm{osc}$ = 0.27 $\pm$ 0.01, and with vertical offset $A_\mathrm{static}$ = 0.008 $\pm$ 0.006. }
    \label{fig:supplramsey}
\end{figure}
\begin{figure}[!h]
    \centering
    \includegraphics{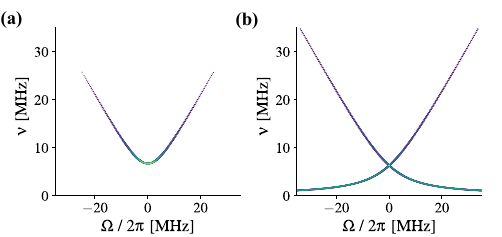}
    \caption{Fourier transform of simulated Rabi oscillations at \textbf{(a)} 82.2 mT and \textbf{(b)} 0 mT versus resonance offset $\Omega$. Both simulations are for $A_\mathrm{iso}$ = 4.5 GHz with a 0.95 mT microwave drive that is in resonance with the $\omega_{34}$ transition at the respective field. The simulation at 82.2 mT shows the dependence that one expects for a two-level system, consisting of a single Rabi oscillation frequency. The zero-field simulation shows two frequency components, as expected in the zero-field limit. The sum and difference frequencies of the two components correspond to $2\nu_\mathrm{eff}$ and $2|\Omega|$, respectively \cite{nishimura_rabi-oscillation_2021}.}
    \label{fig:zrofieldRabi}
\end{figure}

A shorter set of Ramsey fringes, recorded under the same experimental conditions as the Rabi oscillations above, are shown in Fig. \ref{fig:supplramsey}. Figure \ref{fig:zrofieldRabi} shows simulated Rabi oscillation spectra at 82.2 mT and at 0 mT in panels (a) and (b), respectively. While a single Rabi oscillation frequency $\nu_\mathrm{eff}$ is observed in the presence of a magnetic field, two frequencies are observed at zero field.

\section{Supplementary DEMUR data} \label{app:DEMUR}

\begin{figure}[!t]
    \centering
    \includegraphics{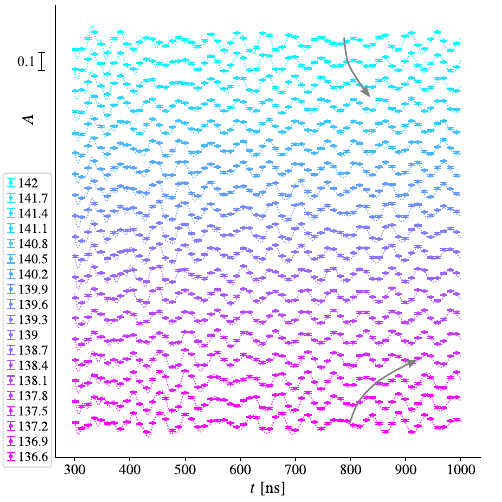}
    \caption{Time-domain transverse field data of Si with 3.9 GHz microwave drive recorded at different magnetic fields (see legend for field in mT), showing binned data (squares) and fits (dashed lines). $t = 0$ corresponds to the arrival of the muon, such that the pulse starts at $t_\mathrm{p} = 300$~ns. Since the fast oscillating component at $\omega_{34}$ decays rapidly, the experimental data were binned for better visualization of the two slower oscillations. Beatings with period around 300 ns between the diamagnetic muons and the lower eigenfrequency $\omega_{12}$ of muonium are readily observable at the wings and central part of the spectrum. The variation of this slow beating period with the magnetic field is mainly due to the field dependence of $\omega_{12}$ (see also gray arrows).}
    \label{fig:allTFs}
\end{figure}

As stated in the main text, the muon frequencies experience strong damping when the single-quantum ESR transitions are driven resonantly. The raw experimental data that clearly show this effect are summarized in Fig. \ref{fig:allTFs}, where the binned data and the corresponding three-component fits are illustrated. The data exhibit an envelope modulation at frequency $\nu_\mathrm{beat}$ on the order of 1/(300 ns). This envelope modulation is due to interference of the diamagnetic muon fraction (from aluminium parts of the resonator and from Si itself) and the $\omega_{12}$ muon precession of bond-centered muonium, whereas the oscillation at $\omega_{34}$ is difficult to observe in the displayed time windows. The variation of the slow envelope modulation with magnetic field reflects the field dependence of $\omega_{12}$: The four lowest curves show a progressive lowering of $\nu_\mathrm{beat}$, which is best recognized by following the second or third modulation maximum around 700 ns and 900 ns, respectively (see also gray arrows). The underlying reason is the reduction in $\omega_{12}$ as the resonance of $\omega_{13}$ is approached from smaller magnetic fields. The corresponding effect for magnetic fields beyond the $\omega_{24}$ resonance can be seen in the three uppermost curves. On resonance, the beating pattern vanishes due to the aforementioned damping of the $\omega_{12}$ (and the $\omega_{34}$) transition. Between the two ESR single-quantum resonances, the beating pattern becomes again clearly visible. However, in this region, it is difficult to infer the field dependence of $\omega_{12}$ just from looking at the beating pattern.

\begin{figure}[!t]
    \centering
    \includegraphics{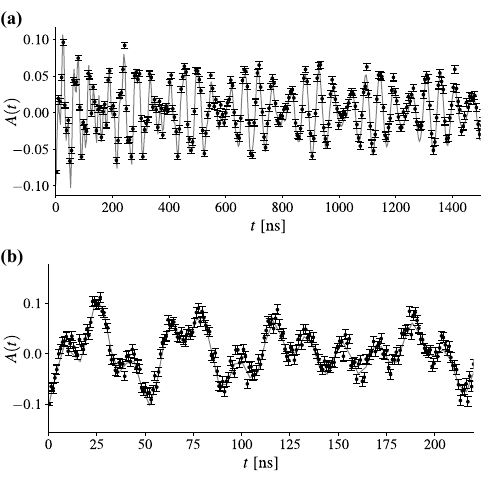}
    \caption{Transverse-field asymmetry measured in Si at a magnetic field of 138.1 mT without microwaves to extract hyperfine parameters $A_\parallel$ and $A_\perp$, showing binned data (circles) and fit (solid gray). \textbf{(a)} a window up to 1500 ns and \textbf{(b)} a zoom on the first 220 ns with the binning used for the actual fit.}
    \label{fig:ABfit}
\end{figure}

As mentioned in the main text, the hyperfine parameters $A_\parallel$ and $A_\perp$ were extracted from TF \textmu{}SR without microwaves, which is shown in Fig. \ref{fig:ABfit}. The data were fitted by three damped oscillations (solid gray). The first component at frequency $\omega_\mu = \gamma_\mu B_0 = 2 \pi \cdot (18.688 \pm 0.0013$~MHz) has an amplitude of $(38.4 \pm 0.5)\cdot 10^{-3}$ and a 
damping of $0.05 \pm 0.008$ \textmu{}s$^{-1}$. The second component at frequency $\omega_{12}$ has an amplitude of (31.3 $\pm$ 0.7)$\cdot 10^{-3}$ and a damping of 0.363 $\pm$ 0.019 \textmu{}s$^{-1}$. The third component at frequency $\omega_{34}$ has an amplitude of (50.0 $\pm$ 1.3)$\cdot 10^{-3}$ and a damping of 2.93 $\pm$ 0.11 \textmu{}s$^{-1}$. The fit parametrized the muon frequencies $\omega_{12}$ and $\omega_{34}$ by the hyperfine parameters and $\omega_\mu$, which yields $A_\parallel / (2\pi)$ = 67.58 $\pm$ 0.08 MHz and $A_\perp / (2\pi)$ = 35.55 $\pm$ 0.07 MHz. Note that there is a significant phase shift between the oscillation components at $\omega_{12}$ and at $\omega_{34}$. For the $\omega_{12}$ component, the oscillation phase of 203 $\pm$ 2$^\circ$ is close to the phase 212 $\pm$ 1$^\circ$ of the diamagnetic $\omega_\mu$ component. However, for the $\omega_{34}$ component, a phase of 134 $\pm$ 1$^\circ$ results. We tentatively ascribe this phase shift to the orientation of the sample and the azimuthal angular offset between the transverse component of the incoming muon, given by the instrument, and the transverse hyperfine component $A_\perp$, given by the (111) crystal axes of the sample.

The DEMUR effects that have been measured in TF mode are in principle also observed in LF mode due to the canted muon states. Basically, the change from TF to LF mode only changes the weighting factors of the oscillations at $\omega_{12}$ and $\omega_{34}$. As an example, we have obtained amplitudes of (37.9 $\pm$ 1.0)$\cdot 10^{-3}$ and (8.0 $\pm$ 1.4)$\cdot 10^{-3}$ for $\omega_{12}$ and $\omega_{34}$ in LF mode at 138.1 mT, which is the same field as for the TF data in Fig. \ref{fig:ABfit}. In this special case, the amplitude of the $\omega_{12}$ component therefore increases when changing from TF to LF mode. However, the $\omega_{34}$ component is very small in LF mode, such that TF mode is better suited for DEMUR. Note that these constellations can be inferred directly from the muon quantization axes plotted in Fig. \ref{fig:silevels2}(b): The $\omega_{34}$ transition is quantized at an angle that is rather close to the $z$-axis, so that the precessing components will be larger in TF mode than in LF mode. The $\omega_{12}$ transition, on the contrary, is quantized at an angle close to 45$^\circ$, resulting in precessing components of similar amplitudes in LF and TF mode.

\newpage

\section{Analytical formulas for DEMUR spectrum} \label{app:analytical}
The analytical expressions for the muon frequencies under microwave irradiation are provided in this Appendix. The expressions are based on diagonalization of the Hamiltonian including the drive term $\hat{\cal{H}}_{1} = \hbar \omega_1 \hat{S}_x$, as described elsewhere \cite{jeschke_generation_1996,schweiger_principles_2001}. The starting point of the procedure is the Hamiltonian
\begin{equation}
	\hat{\cal{H}}_{0} / \hbar = {\omega_S \cdot \hat{S}_z}\  +\  {\omega_I \cdot \hat{I}_z}\  +\  {A_\parallel \cdot\hat{S}_z \hat{I}_z + A_\perp \cdot \hat{S}_z \hat{I}_x} \label{eq:Ham_Si}
\end{equation}
where $A_\parallel$ is the secular (longitudinal) and $A_\perp$ the non-secular (transverse) component of the hyperfine interaction.

For ease of analysis, the Hamiltonian is transformed into a rotating coordinate system by replacing the electron Larmor frequency $\omega_S$ with the resonance offset $\Omega_S$. This Hamiltonian is diagonalized by unitary transformation with
\begin{equation}
    \hat{U}_1 = \exp{-i (\xi \hat{I}_y + \eta 2 \hat{S}_z \hat{I}_y)}. \label{eq:U1_diag}
\end{equation}
The Hamiltonian in Eq.~\eqref{eq:Ham_Si} (in the rotating frame and with $\hbar \rightarrow 1$) then becomes
\begin{equation}
    \hat{\cal{H}}_0^{\ \mathrm{st}} = \Omega_S \hat{S}_z + \frac{\omega_+}{2}\,\hat{I}_z + \omega_-\,\hat{S}_z\hat{I}_z
\end{equation}
where $\omega_\pm$ are muon combination frequencies, i.e. $\omega_\pm = \omega_{12} \pm \omega_{34}$. Likewise, the excitation Hamiltonian $\hat{\cal{H}}_{1}$ in this singly-tilted frame becomes
\begin{equation}
    \hat{\cal{H}}_1^{\ \mathrm{st}} = \omega_1 (\cos{\eta}\,\hat{S}_x + \sin{\eta}\,2\hat{S}_y\hat{I}_y)
\end{equation}
The full Hamiltonian $\hat{\cal{H}}_0^{\ \mathrm{st}} + \hat{\cal{H}}_1^{\ \mathrm{st}}$ is diagonalized in two further steps. First, the single-quantum drive term proportional to $\cos{(\eta)}$ is diagonalized by transformation with\begin{equation}
    \hat{U}_2 = \exp{-i \left((\theta_{13}+\theta_{24})/2 \cdot \hat{S}_y + (\theta_{13}-\theta_{24})\, \hat{S}_y \hat{I}_z \right)}
\end{equation}
with the angles $\theta_{13}$ and $\theta_{24}$ given by $\theta_{ij} = \arctan{\frac{-\omega_1 \cos{\eta}}{\omega_-/2 \pm \Omega_S}}$, respectively. The full Hamiltonian in the doubly-tilted frame becomes
\begin{align}
    \hat{\cal{H}}^{\ \mathrm{dt}} &= \Omega_S^{\ \mathrm{dt}} \,\hat{S}_z + \frac{\omega_+}{2}\,\hat{I}_z + \omega_-^{\ \mathrm{dt}}\,\hat{S}_z\hat{I}_z \nonumber \\ &+ \omega_1 \sin{\eta} \cos{\theta} \, 2\hat{S}_y\hat{I}_y - \omega_1 \sin{\eta} \sin{\theta} \, \hat{I}_x \label{eq:Hdt}
\end{align}
with the difference angle $\theta = (\theta_{13} - \theta_{24})/2$. The eigenfrequencies in this frame are defined by
\begin{equation}
    \Omega_S^{\ \mathrm{dt}} = (\omega_{13}^{\ \mathrm{dt}} + \omega_{24}^{\ \mathrm{dt}})/2, \quad \quad \omega_-^{\ \mathrm{dt}} = \omega_{13}^{\ \mathrm{dt}} - \omega_{24}^{\ \mathrm{dt}}
\end{equation}
with 
\begin{align}
    \omega_{13}^{\ \mathrm{dt}} &= (\Omega_S + \omega_-/2) \cos{\theta_{13}} - \omega_1 \cos{\eta} \sin{\theta_{13}}\\
    \omega_{24}^{\ \mathrm{dt}} &= (\Omega_S - \omega_-/2) \cos{\theta_{24}} - \omega_1 \cos{\eta} \sin{\theta_{24}}
\end{align}
Note that the notation used in these equations is almost identical to the expressions found in Ref.~\cite{schweiger_principles_2001}. The definition of $\omega_{ij}^{\ \mathrm{dt}}$ as well as the proportionality in the Hamiltonian are here, however, different than in Ref.~\cite{schweiger_principles_2001} and formally equivalent to their original definitions found in Ref.~\cite{jeschke_generation_1996}.
\\To diagonalize the off-diagonal terms in the Hamiltonian Eq.~\eqref{eq:Hdt}, only the most relevant off-diagonal term related to zero-quantum coherence is kept, while the others are truncated. In particular, the off-diagonal $\sin{(\theta)}$ term is truncated, while the $\cos{(\theta)}$ term is expanded into zero- and double-quantum coherence terms $\hat{S}_x\hat{I}_x \pm \hat{S}_y\hat{I}_y$, of which only the zero-quantum term is kept \cite{schweiger_principles_2001}.
\\This truncated Hamiltonian in the doubly-tilted frame is diagonalized by a third frame transformation,
\begin{equation}
    \hat{U}_3 = \exp{-i \cdot \chi \left( \hat{S}_x \hat{I}_y  - \hat{S}_y \hat{I}_x \right)}
\end{equation}
with the angle
\begin{equation}
    \chi = \arctan{\frac{\omega_1 \sin{\eta} \cos{\theta}}{\pm\Omega_S^{\ \mathrm{dt}} - \omega_+/2}},
\end{equation}
where the plus or minus sign for $\Omega_S^{\ \mathrm{dt}}$ selects either the zero- or the double-quantum transition, i.e. $\omega_{23}$ or $\omega_{14}$. The truncated Hamiltonian in the triply-tilted frame thus becomes,
\begin{equation}
    \hat{\cal{H}}^{\ \mathrm{tr}} = \Omega_S^{\ \mathrm{tr}} \,\hat{S}_z + \omega_I^{\ \mathrm{tr}}\,\hat{I}_z + \omega_-^{\ \mathrm{dt}}\,\hat{S}_z\hat{I}_z \label{eq:Htr}
\end{equation}
with it's eigenfrequencies,
\begin{align}
    \Omega_S^{\ \mathrm{tr}} &= \pm\Omega_S^{\ \mathrm{dt}} \cos^2{\frac{\chi}{2}} + {\frac{\omega_+}{2}} \sin^2{\frac{\chi}{2}} + {\frac{\omega_1}{2}} \sin{\eta} \cos{\theta} \sin{\chi}\\
    \omega_I^{\ \mathrm{tr}} &=  \pm \Omega_S^{\ \mathrm{dt}} \sin^2{\frac{\chi}{2}} + {\frac{\omega_+}{2}}\cos^2{\frac{\chi}{2}}  - {\frac{\omega_1}{2}} \sin{\eta} \cos{\theta} \sin{\chi} \label{eq:witr}
\end{align}
where the plus and minus signs again apply for either a drive at the zero- or the double-quantum transition.

The Hamiltonian $\hat{\cal{H}}^{\ \mathrm{tr}}$ provides the energy levels and transition frequencies in presence of a resonant microwave drive. Due to the truncation of $\hat{\cal{H}}^{\ \mathrm{dt}}$, the expressions are only valid as long as the truncated elements are negligible. Explicitly, the condition is $\omega_1\sin{\eta} \ll \omega_+$ and the magnitude of $\omega_+$ is on the order of 30 MHz for our conditions. This condition is thus well fulfilled for the TF DEMUR experiments with $\omega_1\sin{\eta}$ on the order of 2.5 MHz. For the LF experiments at full drive, however, the Rabi oscillations on the double-quantum transition were reaching 7 MHz, where contributions from truncated parts become more important.

Formally, the muon frequencies $\omega_{12}^{\ \mathrm{tr}}$ and $\omega_{34}^{\ \mathrm{tr}}$ are obtained directly from the Hamiltonian via $\omega_{ij}^{\ \mathrm{tr}} = \omega_I^{\ \mathrm{tr}} \pm \omega_-^{\ \mathrm{dt}}/2$, where $\omega_{12}^{\ \mathrm{tr}}$ is the sum and $\omega_{34}^{\ \mathrm{tr}}$ the difference. The DEMUR spectrum is obtained by tracing the dependence of these frequencies on the resonance offset $\Omega_S$. However, there are two critical points to consider with respect to $\omega_I^{\ \mathrm{tr}}(\Omega)$. First, there are two different possibilities of $\omega_I^{\ \mathrm{tr}}$ that differ significantly in the vicinity of double- and zero-quantum transitions. To include effects related to both transitions, we used $\widetilde{\omega_I}^{\ \mathrm{tr}} = \omega_I^{\ \mathrm{tr, +}} + \omega_I^{\ \mathrm{tr, -}} - \omega_+/2$, where the plus and minus superscripts denote the corresponding choice of the sign of $\Omega_S^{\ \mathrm{dt}}$ in Eq.~\eqref{eq:witr}. Second, there is an avoided crossing of $\omega_I^{\ \mathrm{tr}}$ and $\Omega_S^{\ \mathrm{tr}}$ at either the zero- or double-quantum transition, which corresponds to the Rabi splitting of the transition. When passing beyond either $\omega_{14}$ or $\omega_{23}$, the muon splitting $\omega_I^{\ \mathrm{tr}}$ must be substituted by the resonance offset $\Omega_S^{\ \mathrm{tr}}$. The locations of the respective resonance crossings were determined from the condition $\chi = -\pi/2$. Note that the shift of the double-quantum transition $\omega_{14}$ indicated by the dashed gray curve in Fig. \ref{fig:LFdrive14}(c) has also been obtained from the above condition on $\chi$. 

Besides these particularities related to $\omega_I^{\ \mathrm{tr}}(\Omega_S)$ at multi-quantum transitions, the Rabi splittings in the single-quantum transitions are directly contained in $\omega_-^{\ \mathrm{dt}}$. Also note that in absence of hyperfine anisotropy, e.g. for $B=0$, the muon frequencies $\omega_{12}^{\ \mathrm{dt}}$ and $\omega_{34}^{\ \mathrm{dt}}$ related to the Hamiltonian $\hat{\cal{H}}^{\ \mathrm{dt}}$ fully contain the DEMUR spectrum. In this case, the frequencies directly follow from $\omega_{ij}^{\ \mathrm{dt}} = (\omega_+ \pm \omega_-^{\ \mathrm{dt}})/2$.

The muon eigenfrequencies obtained in this way constitute the principal muon oscillation frequency. At microwave resonances, however, muon frequencies become split by the Rabi frequency. The resultant dynamics require a more involved analysis. For a TF DEMUR experiment, for instance, the procedure is as follows: The initial state $\sigma_0$ = $\hat{I}_x$ of the experiment in the laboratory frame is transformed into the triply-tilted interaction frame. Subsequently, the microwave drive is applied by unitary evolution under $\hat{\cal{H}}^{\ \mathrm{tr}}$ during time $t$. The final state is transformed back into the laboratory frame and the time-dependent observable $<\hat{I}_x>(t)$ is extracted. Using a dedicated library for the software Mathematica\textsuperscript{\textregistered}\cite{spinop}, the resultant analytical expressions can be computed. While these formulas are too long to be shown here, the dynamics are readily visualized in spectral domain upon (numerical) Fourier transform of $<\hat{I}_x>(t)$.

\begin{figure}[t!]
	\centering
	\includegraphics{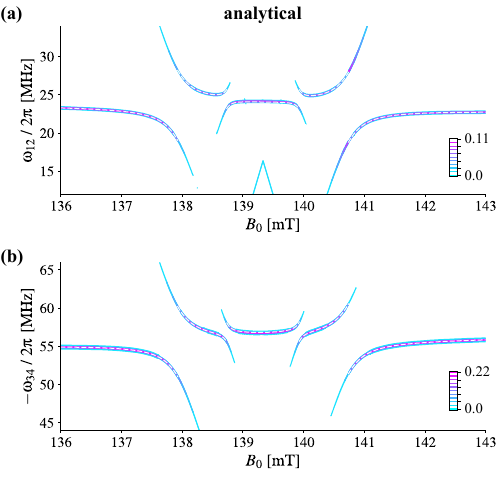}
	\caption{TF muon frequencies \textbf{(a)} $\omega_{12}$  and \textbf{(b)} inverted $\omega_{34}$. The contour levels show Fourier spectra of time traces $<\hat{I}_x>(t)$, which correspond to the transverse muon signal with a microwave drive. The time traces were calculated using analytical expressions derived as described in the text, which are parametrized by $\Omega_S$, $\omega_I$, $A_\parallel$, $A_\perp$ and $\omega_1$. The resonance offset $\Omega_S$ and the muon Larmor frequency $\omega_I$ were calculated from the magnetic field, using $g_e$ = 1.9999 and $\omega_\mathrm{uw} / 2\pi$ = 3.9 GHz for the electron. The other parameters were as in Fig. \ref{fig:TFDEMUR} in the main text. In addition to the FT contours, the analytical muon eigenfrequencies are superimposed as white dashed lines. These show excellent agreement with the dominant Fourier contours, except for subtle differences at the center of the spectra that are discussed in the text.
	}
	\label{fig:TFDEMUR_pattern}
\end{figure}

Fig. \ref{fig:TFDEMUR_pattern} shows contour levels of the Fourier spectra around $\omega_{12}$ in panel (a) and around $-\omega_{34}$ in panel (b). The abscissa is here the magnetic field $B_0$, which is proportional to the resonance offset $\Omega_S$ that is used for the calculations. The transient Rabi splittings due to avoided crossings are clearly seen in the figures. In addition, the muon eigenfrequencies obtained as described above are superimposed as white dashed lines, which are best identified around the discontinuities on resonance. These superimpose very well onto the spectra. On a very close look, one can infer a slight offset in both $\omega_{12}$ and $\omega_{34}$ at the center of the spectra (around 139.4 mT). These are related to the sign of $\Omega_S^{\ \mathrm{dt}}$ in the eigenfrequencies of $\hat{\cal{H}}^{\ \mathrm{tr}}$. For the analytical expressions underlying the FT spectra, it is only possible to make one particular choice of the sign. For the muon eigenfrequencies, on the contrary, off-resonant shifts due to excitation of both zero- and double-quantum transitions are added together at the center (see $\widetilde{\omega_I}^{\ \mathrm{tr}}$ above). Accordingly, the analytical muon eigenfrequencies have a more pronounced shift than the FT spectra. Along the same lines, there is an extra feature observed in the $\omega_{12}$ spectra below 15 MHz. This feature is related to Rabi oscillations in the static muon polarization, as driven experimentally in LF mode. The abrupt kink at the center is again related to the choice of the sign of $\Omega_S^{\ \mathrm{dt}}$.

\begin{figure}[t!]
    \centering
    \includegraphics{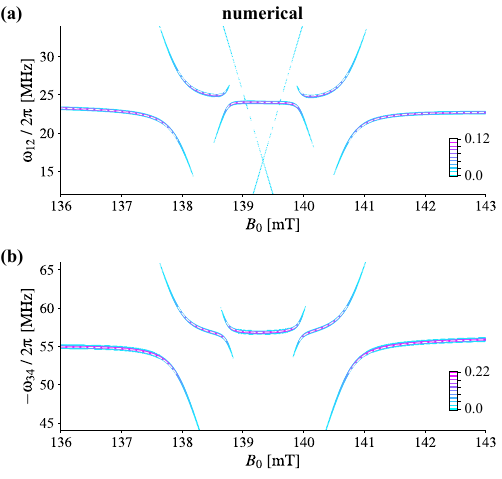}
    \caption{TF muon frequencies \textbf{(a)} $\omega_{12}$  and \textbf{(b)} inverted $\omega_{34}$. The contour levels show Fourier spectra of time traces $<\hat{I}_x>(t)$, which correspond to the transverse muon signal with a microwave drive. The time traces were calculated numerically using Spidyan, with exactly the same parameters as used in Fig. \ref{fig:TFDEMUR_pattern}. In addition to the FT contours, the analytical muon eigenfrequencies are superimposed as white dashed lines. These virtually superimpose with the dominant Fourier contours and approve consistency among the used modeling methods.
    }
    \label{fig:TFDEMUR_pattern_spidyan}
\end{figure}

In order to verify the analytical calculations above, Fourier spectra obtained by numerical simulation are shown in Fig. \ref{fig:TFDEMUR_pattern_spidyan}. The simulations use exactly the same parameters as for Fig. \ref{fig:TFDEMUR_pattern}. The numerical Fourier contours agree well with the muon eigenfrequencies overlaid as white dashed lines, which confirms consistency between the two approaches. Note that the kink feature in $\omega_{12}$ discussed in the previous paragraph is here extended and shows the expected crossing of the multi-quantum transition's off-resonant branches.

For the experimental data in this study, the analytical muon eigenfrequencies were sufficient for data analysis. This is due to the fast damping on resonance, which inhibited the formation of sustained Rabi splittings in our TF DEMUR experiments. In the absence of fast damping, such Rabi splittings would be observable and direct fitting of time-domain data to analytical expressions could be envisaged to retrieve the parameters.

During the fitting based on muon eigenfrequencies in this study, the discontinuities in the eigenfrequencies posed a complication. In fact, data points that were very close to resonance \emph{pinned} the $\chi^2$ minimum and resulted in discontinuities in $\chi^2$ maps that sampled the relevant range of $\omega_1$ and $g_e$. As shown in Fig. \ref{fig:TFDEMUR}, such data points have therefore been excluded for the fitting procedure. In this way, the parameters and their errors could be extracted from well-defined and continuous $\chi^2$ maps. In view of the fast on-resonant damping, such a procedure is also reasonable with respect to the actual ESR line shape. Basically, the resonant part of the DEMUR spectrum is most sensitive to the ESR line shape, which generally exhibits homogeneous and inhomogeneous broadening. By excluding the points on resonance, one also avoids eventual changes in muon frequencies related to line broadening that are not contained in the analytical muon eigenfrequencies. At a resonance offset beyond the ESR line width, on the contrary, the influence of the ESR line shape is less important.

\begin{figure}
    \centering
    \includegraphics{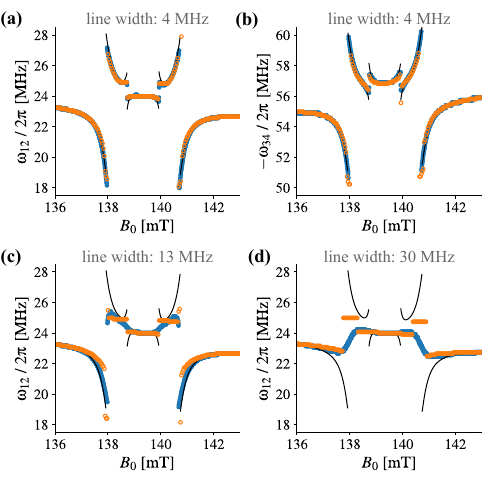}
    \caption{Muon frequencies $\omega_{12}$ in panels \textbf{(a, c, d)} and inverted $\omega_{34}$ in panel \textbf{(b)}. The black lines are the analytically calculated eigenfrequencies. The colored points correspond to the maximum of Fourier spectra computed from simulated $<\hat{I}_x>(t)$ time traces. The filled blue circles originate from the homogeneously broadened line, while the open orange circles originate from the inhomogeneously broadened line. For panels (a) and (b), the homogeneous and inhomogeneous line widths were 4.2 MHz, corresponding to $T_2^e$ = 75 ns for the homogeneous case. For panels (c) and (d), the homogeneous and inhomogeneous line width were 12.7 MHz and 31.2 MHz, respectively. For the homogeneous case, these correspond to $T_2^e$ times of 25 ns and 10 ns, respectively.
    }
    \label{fig:TFDEMUR_pattern_damping}
\end{figure}

In order to quantify the influence of the ESR line width for the parameters in this study, we compared numerical simulations with homogeneous or inhomogeneous broadening to analytically calculated muon eigenfrequencies. For homogeneous line broadening, a relaxation time $T_2^e$ = 75 ns was assumed, which corresponds to the experimental Ramsey decay time in Fig. \ref{fig:LFdrive14}. For inhomogeneous line broadening, the time-domain traces $<\hat{I}_x>(t)$ were simulated for a Gaussian ESR line with FWHM equal to the homogeneously broadened line. Accordingly, both line shapes under consideration had a FWHM of 4.2 MHz. The results are summarized in Fig. \ref{fig:TFDEMUR_pattern_damping}(a) and (b) for $\omega_{12}$ and $\omega_{34}$, respectively. The black lines are the analytically calculated eigenfrequencies. The colored points correspond to the maximum of Fourier spectra computed from simulated $<\hat{I}_x>(t)$ time traces. The shaded blue circles originate from the homogeneously broadened line, while the open orange circles originate from the inhomogeneously broadened line. The latter has less field points, because each field point requires averaging over the inhomogeneously broadened line.

As is readily seen in Fig. \ref{fig:TFDEMUR_pattern_damping}(a) and (b), there is no significant influence of the ESR line width on the muon frequencies. For reference, the same analysis was performed for larger line widths of 12.7 MHz and 31.2 MHz, with the resulting $\omega_{12}$ shown in Fig. \ref{fig:TFDEMUR_pattern_damping}(c) and (d), respectively. For these larger line widths, the effects on $\omega_{12}$ are more pronounced. For homogeneous line broadening, the Rabi splittings become smeared out, since the Rabi oscillations are strongly over damped. For inhomogeneous line broadening, a step pattern emerges. Therefore, for cases where the ESR line width approach these values, the fitting procedure has to be adapted.

\section{Supplement for Ramsey fringes and Rabi oscillations with double-quantum drive}  \label{app:DQ}

\subsection{Experimental optimization of Rabi oscillations}  \label{app:Rabi}

\begin{figure}[!h]
    \centering
    \includegraphics{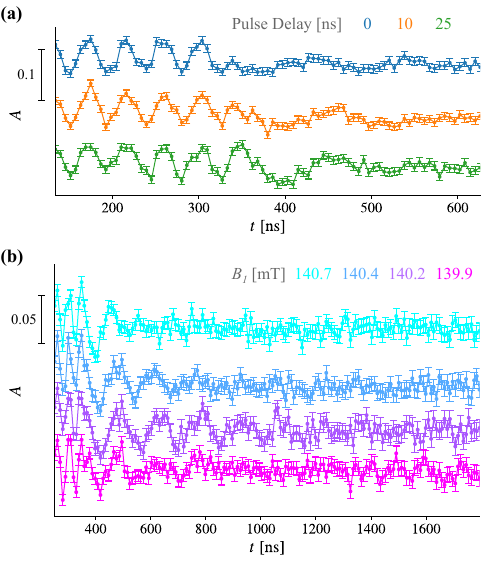}
    \caption{Experimental optimization of Rabi oscillations in LF geometry at the $\omega_{14}$ transition. \textbf{(a)} Muon asymmetry $A(t)$ for three different delays of the 3.9 GHz microwave pulse with respect to its shortest possible delay (see legend) in an applied  field of $B_0$ = 140.7 mT. \textbf{(b)} Muon asymmetry $A(t)$ using the optimized pulse delay of 25 ns at four different magnetic fields (see legend). The magnetic field values were extracted from residual transverse oscillations related to muons stopped inside aluminium. The solid lines are guides to the eye.
    }
    \label{fig:LFDQ_raw}
\end{figure}

The observation of sustained Rabi oscillations in Fig. \ref{fig:LFdrive14}(a) required a number of optimization steps, which are detailed in the following. First, we noticed a rather pronounced dependence of the Rabi oscillation amplitude on the arrival time $t_\mathrm{p}$ of the microwave pulse. The relevant data are shown in Fig. \ref{fig:LFDQ_raw}(a), where three different pulse delay settings are shown. For our initial choice of zero added delay (blue), the Rabi oscillation amplitude was comparably small. A much larger oscillation amplitude was achieved for an added delay of 25 ns (green). In principle, such a delay dependence is reminiscent of well-known \emph{blind spot} effects in ESR pulse sequences for electron-nuclear coherence transfer by a block of two $\pi/2$ pulses \cite{schweiger_principles_2001}. 

\begin{figure}[!t]
	\centering
	\includegraphics{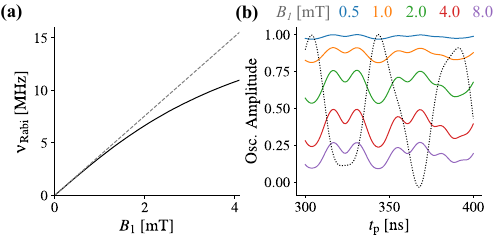}
	\caption{Supplementary simulations of Rabi oscillations. \textbf{(a)} Dependence of apparent Rabi frequency $\nu_\mathrm{Rabi}$ as a function of driving field $B_1$ (solid). The parameters for the Hamiltonian were (in units of Hz) $\omega_S$ = 3.9 GHz, $\omega_I$ = 18.9 MHz, $A_\parallel$ = 67.6 MHz, $A_\perp$ = 35.6 MHz, $\omega_\mathrm{uw}$ = $\omega_{14}$. The dashed curve is the linear dependence predicted from the transition moment $\gamma_\mathrm{e} \sin{\eta}$. \textbf{(b)} Solid lines: Amplitude of Rabi oscillations in muon polarization due to drive at $\omega_{23}$ versus pulse delay $t_\mathrm{p}$ after muonium formation for various driving amplitudes $B_1$ (see legend for colors). Other than the swept variables $B_1$ and $t_\mathrm{p}$, the simulation parameters were exactly as for Fig. \ref{fig:LFdrive14}(c). Dotted line: Normalized muon polarization at the time when the pulse is applied, showing the characteristic beating pattern.}
	\label{fig:silevels_suppl}
\end{figure}

To understand the origin of the pulse delay dependence in our experiments, we have simulated Rabi oscillations as a function of the pulse delay $t_\mathrm{p}$. In the simulation, the microwave drive was applied at the unperturbed resonance frequency of the double-quantum transition. The simulation employed different amplitudes of the driving field $B_1$. The dependence of the Rabi frequency $\nu_\mathrm{Rabi}$ on $B_1$ simulated in this way is shown in by the black curve in Fig.~\ref{fig:silevels_suppl}(a). The dashed curve is the linear dependence expected from the transition moment $\gamma_{14}$. As expected, the linear dependence coincides with the simulation in the limit of small driving amplitudes. The variation of the Rabi oscillation amplitude with the pulse delay $t_\mathrm{p}$ is shown in Fig. \ref{fig:silevels_suppl}(b) for different driving amplitudes. With increasing driving strength, the oscillation amplitude is reduced and becomes dependent on $t_\mathrm{p}$. The muon polarization just before application of the pulse is shown by the dashed line and illustrates the characteristic muon beating pattern in LF mode.

The reduction of the oscillation amplitude and $t_\mathrm{p}$-dependence might also be partly related to the drive-induced shift of the $\omega_{14}$ transition [see Fig. \ref{fig:LFdrive14}(c)]. In any case, it is unlikely that the correct magnetic field and pulse delay are found without experimental optimization. As an example, the data in Fig. \ref{fig:LFDQ_raw}(a) were recorded in a magnetic field of 140.7~mT. A subsequent optimization of the magnetic field revealed that this was not the appropriate setting, as shown in Fig. \ref{fig:LFDQ_raw}(b). As is directly evident from the raw data, the magnetic field was set to 140.2 mT for the experiments presented in the main text in Fig. \ref{fig:LFdrive14}.

\subsection{Ramsey fringes: Phase cycle and simulations}  \label{app:Ramsey}

\begin{figure}[!h]
    \centering
    \includegraphics{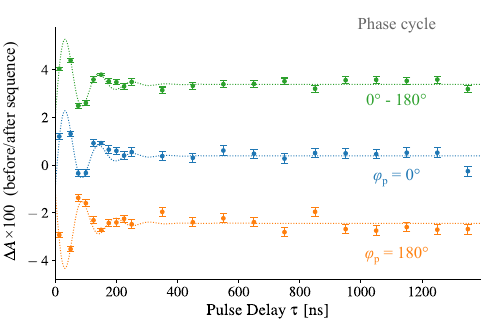}
    \caption{The asymmetry difference, $\Delta A$ as a function of delay, $\tau$ measured in a field of 140.2~mT using different phases in a Ramsey experiment. The data and corresponding fit shown in green are the same as in Fig. \ref{fig:LFdrive14} for the full range of $\tau$. The data points in blue and orange show the experimental data for a phase offset of the second pulse by 0$^\circ$ and 180$^\circ$ with respect to the first pulse, respectively. The dashed lines are fits to the combined signal adjusted to the phase constellation. The time window before the sequence spans from 2 ns to 332 ns, while the window after the sequence starts 20 ns after second pulse and stops at 4 \textmu{}s. For illustration purposes, $\Delta A$ is vertically offset by 0.03. The orange curve corresponds to the raw data without offset and thus directly indicates how the mean polarization of 0.085 before the pulse sequence is altered.
    }
    \label{fig:LFDQ_pcyc}
\end{figure}

\begin{figure}[!h]
    \centering
    \includegraphics{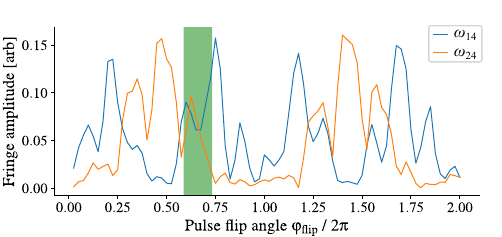}
    \caption{Simulation of Ramsey fringes related to the double-quantum transition $\omega_{14}$ (blue) and the adjacent single-quantum transition $\omega_{24}$ (orange). The simulation is based on two pulses at delay $\tau$ with flip angle on  as indicated on the abscissa. The amplitudes on the ordinate correspond to the (magnitude) Fourier coefficient of the simulated $\tau$-dependent Ramsey fringes at the corresponding detuning $\Omega = \omega_{ij} - \omega_\mathrm{uw}$, where $\tau$ was incremented from 0 to 1000 ns in 2 ns steps. The shaded green area denotes the flip angle range that represents the pulses used in the experiment. 
    }
    \label{fig:Ramsey_sim}
\end{figure}

As mentioned in the main text, Ramsey fringes on bond-centered muonium employed a two-step phase cycle on the second microwave pulse. Figure \ref{fig:LFDQ_pcyc} shows the change in asymmetry $\Delta A$ for phase offsets of 0$^\circ$ (blue) and 180$^\circ$ (orange) with respect to the first pulse. The combined data formed by taking the difference are shown in green. The fit to experimental data shown by the dashed lines is based on these combined data. The fitting parameters as well as the initial part of the combined data are shown in Fig. \ref{fig:LFdrive14}(b) in the main text. The advantage of this phase cycle is that the relevant change in asymmetry $\Delta A$ can extracted by difference formation. It is advisable to further extend the two-step phase cycle to a four-step phase cycle with a phase increment of 90$^\circ$. Such a four-step cycle will reveal the sign of the resonance offset $\Omega$. The sign of $\Omega$ is particularly relevant to identify the ESR transition that is responsible for the observed Ramsey fringe pattern.

For the four-level system encountered in this study, assignment of the underlying ESR transitions from the available Ramsey data set is actually not a trivial task. The main complication is that an excitation pulse which is resonant with the double-quantum transition does not only generate coherence at the \emph{resonant} $\omega_{14}$ transition, but also on the adjacent $\omega_{24}$ transition. One therefore needs to evaluate to which extent each of the transitions contributes to the resultant Ramsey fringes. Numerical simulations shows that the contributions of either single- or double-quantum evolution during the free evolution period $\tau$, critically depend on the duration of the pulses and the arrival time $t_\mathrm{p}$ of the first pulse upon muon implantation. 

A simulation that mimics our experimental conditions was used to evaluate the fringe amplitude due to evolution on the $\omega_{14}$ or $\omega_{24}$ transition as a function of the pulse flip angle $\varphi_\mathrm{flip}$ on $\omega_{14}$, as shown in Fig. \ref{fig:Ramsey_sim}. The pulse flip angle $\varphi_\mathrm{flip}$ was varied by adjusting the duration of the two excitation pulses. Maximum fringe amplitude corresponds to maximum transfer of muon polarization to electron spin coherence, and vice versa. For the fringes at $\omega_{14}$, the amplitudes are maximized for excitation pulses that realize a flip angle $\varphi_\mathrm{flip}$ that is a multiple of $\pi/2$. For the fringes at $\omega_{24}$, the amplitudes are maximized for pulse flip angles of $\pi$ and $3\pi$.
The shaded green area corresponds to the range of flip angles implemented experimentally. This range has been determined by comparison of experimental Ramsey traces to the Rabi oscillations showed in Fig. \ref{fig:LFdrive14}(a). In particular, averaging of raw Ramsey data sets with $\tau >$ 500 ns provided the muon polarization after the first pulse with very high quality. The flip angle was then estimated based on the polarization after the first pulse and the fitting function of the Rabi oscillations. This procedure results in a range of possible flip angles, since the polarization after the first pulse oscillates. The width of the indicated flip angle range corresponds to the peak-to-peak oscillation amplitude, while the center corresponds to the mean polarization after the first pulse. Within this area, either (i) dominant double-quantum evolution or (ii) a mixture of single- and double-quantum evolution is possible. Other than the flip angle $\varphi_\mathrm{flip}$, also the pulse delay $t_\mathrm{p}$ is relevant. For the depicted simulation, the delay was set to 360 ns, which is based on the two-zone fitting procedure in Fig. \ref{fig:LFdrive14}(a). Since neither the simulation nor the two-zone fitting procedure account for the non-zero rise time of the pulse flanks, we consider this delay setting as an upper limit. For smaller delay settings down to 350 ns, the evolution on the double-quantum transition becomes dominant over the entire range of possible flip angles. We therefore deduce from the simulations that the dominant double-quantum evolution is the most likely scenario in the experiment.

Further evidence for double-quantum evolution is obtained when considering the $g$-factor that results for a particular assignment of the Ramsey fringes. For dominant double-quantum evolution during $\tau$, the experimental Ramsey frequency $\Omega$ is reproduced with a $g$-factor of 2.0012. For equipartition of single- and double quantum evolution, $\Omega$ would result at a $g$-factor of 1.9928 via the combination frequency. Moreover, dominant single-quantum evolution would suggest a $g$-factor of 2.0037. Due to proximity to the more precise $g$-value extracted from TF DEMUR experiments, we deduce that the experimental Ramsey fringes originate principally from coherent evolution on the double-quantum transition $\omega_{14}$.

Given the critical influence of the flip angle $\varphi_\mathrm{flip}$ on the Ramsey fringes, it is worth to mention an important experimental aspect related to its calibration. In the experiments performed here, the pulse durations were set to 72 ns. With the experimental Rabi oscillation period on the order of 147 ns, one might expect a nominal flip angle of $\pi$. However, a short pulse has both rising and falling pulse flanks, whereas the Rabi oscillations are observed throughout the course of a long pulse. The falling pulse flank will thus result in an extra rotation of the spins that is not contained in the Rabi oscillations. The range of possible pulse flip angles deduced from experimental data is therefore beyond $\pi$ (green shaded area in Fig. \ref{fig:Ramsey_sim}). Note that this subtle experimental aspect is due to the technique's unique availability to directly observe driven time-domain dynamics via the muon decay asymmetry. In conventional pulsed magnetic resonance techniques, driven time-domain dynamics are often not observed directly, but rather indirectly by a time increment of the drive. In these indirect schemes, the drive always contains both rising and falling pulse edges. An example are the Rabi oscillations during ESR benchmarking described in Appendix \ref{app:ESR}.

\subsection{Narrowing of inhomogeneous broadening via Rabi oscillations} \label{app:narrowing}

\begin{figure}[!h]
    \centering
    \includegraphics{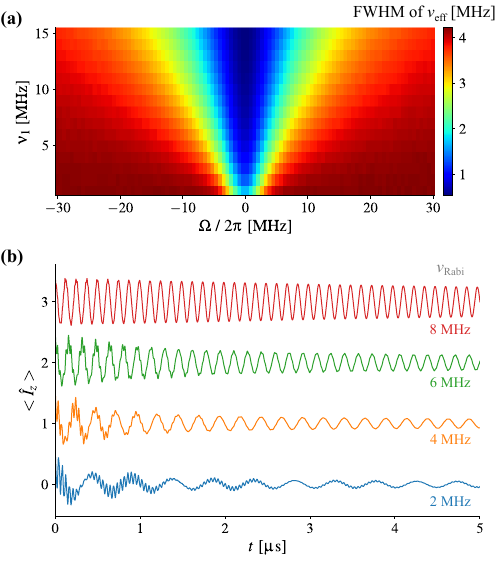}
    \caption{Narrowing of inhomogeneous line under resonant drive. \textbf{(a)} The FWHM of the distribution of $\nu_\mathrm{eff} = (\nu_1^2 + (\Omega/2\pi)^2)^\frac{1}{2}$ as a function of the mean values of $\nu_1$ and $\Omega$. The standard deviation in $\Omega$ was chosen such that the FWHM of the distribution is 4.2 MHz. For $\nu_1$, the FWHM of the distribution was set to 0.4 MHz. \textbf{(b)} Simulated Rabi oscillations for resonant drive at $\omega_{14}$ and Gaussian line broadening with FWHM of 4.2 MHz. The curves are arranged vertically according to the indicated Rabi oscillation frequencies $\nu_\mathrm{Rabi}$. The simulations had the same parameters as used for Fig. \ref{fig:LFdrive14}(c) and incorporated the results calculated therein: At each driving field $B_1$, the magnetic field $B_0$ was set to achieve resonance at $\omega_{14} = 3.9$~ GHz. 
    }
    \label{fig:LFRabi_inhom}
\end{figure}

As stated in the main text, a drive of sufficient strength can narrow both homogeneous and inhomogeneous contributions to line broadening. Homogeneous line narrowing is achieved by CW dynamical decoupling of the muonium center from the bath of distant and nearly-degenerate nuclear spins. Numerical analysis of this many-body mechanism requires cluster-correlation expansion techniques \cite{witzel_multiple-pulse_2007, RenBao_CCE} that are far beyond the scope of the present study. In contrast, narrowing of inhomogeneous contributions can be analyzed by relatively simple means.

The principal reason for drive-induced inhomogeneous line narrowing is the non-linear dependence of the effective Rabi frequency $\omega_\mathrm{eff}$ on the resonance offset $\Omega$ and on the driving strength $\omega_1$. In loose terms, the distribution of $\omega_\mathrm{eff}$ adopts to the (inhomogeneous) distribution of the dominant factor. For large $\Omega$, the main source of Rabi damping is the inhomogeneous line width, while on resonance it is governed by the inhomogeneity of the driving field $\omega_1$. A graphical illustration is provided in Fig. \ref{fig:LFRabi_inhom}(a), where the FWHM of $\nu_\mathrm{eff}$ is shown as a function of $\Omega /2\pi$ and $\nu_1$. The resultant 2D map reflects the limiting cases mentioned above. In the red areas at large offset, $\nu_\mathrm{eff}$ adopts the FWHM of 4.2 MHz that has been input for $\Omega$. In contrast, for the blue areas at large $\nu_1$, $\nu_\mathrm{eff}$ has a narrower distribution, since a FWHM of 0.4 MHz has been input for $\nu_1$.

To confirm that this narrowing mechanism applies to our four-level system with double-quantum drive, we have performed numerical simulations. For feasible simulation times, we have only considered inhomogeneity in $\Omega$ and assumed a fixed value for $\nu_1$. The obtained Rabi oscillations are shown in Fig. \ref{fig:LFRabi_inhom}(b), where the vertically displaced curves correspond to different Rabi oscillation frequencies $\nu_\mathrm{Rabi}$ on the double-quantum transition. As is readily seen, the Rabi damping decays progressively with $\nu_\mathrm{Rabi}$.

Note that we do not expect such an efficient inhomogeneous narrowing for TF DEMUR experiments. The reason is the anti-symmetry of the underlying TF DEMUR spectrum around resonance. This anti-symmetry results from linear contributions of $\Omega$ to the respective spectrum around resonance. The Rabi spectrum, on the contrary, is symmetric around resonance due to the non-linear dependence on $\Omega$ [see also Fig. \ref{fig:zrofieldRabi}(a)].

\bibliography{Mu-uwave}

\end{document}